\documentclass[11pt]{article}
\usepackage[dvips]{graphicx}
\usepackage{amssymb}
\usepackage{amsmath}
\usepackage{amscd}
\usepackage{latexsym}

\catcode`@=11
\renewcommand{\section}{\@startsection{section}{1}{0pt}{\medskipamount}
{\medskipamount}{\large\bf}}
\numberwithin{equation}{section}
\catcode`@=12
\def\a{\alpha}
\def\b{\beta}
\def\g{\gamma}

\def\de{\delta}

\def\Th{\Theta}

\def\r{\rho}

\def\ome{\omega}
\def\Om{\Omega}
\def\La{\Lambda}

\def\1{\bar 1}
\def\2{\bar 2}
\def\3{\bar 3}
\def\4{\bar 4}

\def\eps{\epsilon}
\def\lrc{\,\lrcorner\,}

\newcommand{\C}{\mathbb C}
\newcommand{\R}{\mathbb R}

\newcommand{\Acal}{{\cal A}}

\newcommand{\Fcal}{{\cal F}}

\newcommand{\T}{{\cal T}}

\def\im{\mbox{i}}
\def\N2{$N{=}2$}
\def\pa{\mbox{$\partial$}}
\def\diff{\mbox{d}}

\def\sfrac#1#2{{\textstyle\frac{#1}{#2}}}
\def\>{\rangle}
\def\<{\langle}
\def\+{\dagger}
\def\={\ =\ }
\def\und{\qquad\textrm{and}\qquad}
\def\and{\quad\textrm{and}\quad}
\def\for{\quad\textrm{for}\quad}

\begin{document}
\begin{titlepage}
\setcounter{page}{0}
\begin{flushright}
.
\end{flushright}

\vskip 2.5cm

\begin{center}

{\large\bf Yang-Mills fields in flux compactifications on homogeneous manifolds with SU(4)-structure}
\\[2mm]

\vspace{12mm}
{\large Derek Harland$^\ast$ and Alexander D. Popov$^\dagger$}
\\[2mm]
\noindent {$^\ast$\em Department of Mathematical Sciences, Durham University\\
Science Laboratories, South Road, Durham, DH1 3LE, UK}
\\
{Email: {\tt d.g.harland@durham.ac.uk}}
%\vspace{12mm}
\\[2mm]
\noindent {$^\dagger$\em Bogoliubov Laboratory of Theoretical Physics, JINR\\
141980 Dubna, Moscow Region, Russia}
\\
{Email: {\tt popov@theor.jinr.ru}}
\vspace{12mm}

\begin{abstract}
\noindent
The Spin(7)-instanton equations are natural BPS equations for instantons on 8-manifolds.  We study these equations on nearly K\"ahler and Calabi-Yau torsion manifolds of the form $M\times G/H$, with $G/H$ a coset space and $M$ a product of a torus with Euclidean space.  By imposing $G$-invariance the instanton equations reduce to interesting equations on $M$; for example, equations used by Kapustin and Witten in the geometric Langlands program arise in this way.  We carry out reductions in a number of examples, and where possible present simple solutions.
\end{abstract}
\end{center}

\end{titlepage}

\section{Introduction}

%\noindent
%First paragraph

%\noindent
%{\bf Subsection}. Text continues...

Generalisations of the four-dimensional anti-self-dual Yang-Mills equations to dimensions greater than four were first written down by Corrigan et. al. \cite{Corrigan:1982th}.  The mathematical study of these equations was spurred by the article \cite{DT} of Donaldson and Thomas, which sought high-dimensional analogues of low-dimensional gauge-theoretical topological invariants.  In physics, these equations turned out to be naturally BPS equations, and have in particular become important in compactifications of string, M-, and F-theory.  They have also been used to construct topological field theories \cite{Bau}, whose dimensional reductions have found application both in the geometric Langlands program \cite{Kapustin:2006pk} and an analytic continuation of Chern-Simons theory \cite{Witten:2010cx}.

The most interesting dimensions seem to be 6, 7, and 8, and the corresponding generalised anti-self-dual equations are respectively called the Hermitian-Yang-Mills or Donaldson-Uhlenbeck-Yau equations \cite{DUY}, the $G_2$-instanton equations \cite{SaEarp}, and the Spin(7)-instanton equations \cite{Lewis}.  The first solutions were written down in Euclidean space \cite{WFN,FNIP}, but later it was realised that the equations make sense on any manifold, provided it is equipped with a $G$-structure for a suitable Lie group $G$.  One requires $G={\rm SU}(3)$ in dimension 6, $G=G_2$ in dimension 7, and $G={\rm Spin}(7)$ in dimension 8.  The choice $G={\rm SU}(4)$ is also suitable on an 8-dimensional almost-complex manifold, because an SU(4)-structure lifts to a Spin(7)-structure; the Spin(7)-instanton equations are sometimes called the SU(4)-instanton equations in this context.  For a recent discussion and more references, see for example \cite{DS,P1,HILP}.

Most work on the generalised anti-self-dual Yang-Mills equations has restricted attention to integrable $G$-structures, that is, Riemannian manifolds with holonomy group $G$.  However, if one is interested in string compactifications with fluxes \cite{flux} one should consider instead non-integrable $G$-structures.  The torsion of the $G$-structure, which measures the failure to be integrable, is in this context identified with a 3-form field in supergravity.

With this in mind, we have undertaken to study the Spin(7)-instanton equations on 8-dimensional coset spaces with non-integrable $G$-structures.  We have restricted most of our attention to nearly K\"ahler and Calabi-Yau torsion coset spaces (both of which admit an almost complex structure).  Mathematically, these are quite natural choices: they are obtained by requiring that the torsion 3-form is the real part of a (3,0)-form or a (2,1)-form respectively.  Physically, nearly K\"ahler 6-manifolds are suitable for compactification of both Romans massive supergravity \cite{Lust:2004ig} and heterotic supergravity \cite{CMPZ} to $AdS_4$, while the Calabi-Yau torsion condition is a necessary (though not sufficient) condition in compactifications of heterotic supergravity to 4-dimensional Minkowski space \cite{Str,Lopes Cardoso:2002hd}.

An outline of the rest of this article is as follows.  Section \ref{sec:2} contains an introduction to ${\rm SU}(n)$-structures on $2n$-manifolds, and includes a digression on torsion classes in the case $n=3$.  In section \ref{sec:3} we introduce the Spin(7)-instanton equations, highlighting their relations with the torsionful Yang-Mills equations and with topological quantum field theories.  In section \ref{sec:4} we describe how to construct $G$-invariant gauge fields on coset spaces $G/H$.  We then discuss in sections \ref{sec:5}-\ref{sec:8} the Spin(7)-instanton equations on many examples of compact 8-manifolds of the form $T^d\times G/H$.  In each example, we first impose $G$-invariance on the gauge field, reducing the 8-dimensional Spin(7)-instanton equations to equations on the torus $T^d$.  Where possible, these $d$-dimensional equation are solved.  Besides solutions on the torus, solutions on decompactified spaces $\R^{d-p}\times T^p$ are also considered.  We summarise our results and make some closing comments in section \ref{sec:9}.

\bigskip

\section{Manifolds with ${\rm SU}(n)$-structure}
\label{sec:2}

\subsection{Calabi-Yau torsion and nearly K\"ahler manifolds}

A manifold with $H$-structure is by definition an $m$-dimensional manifold $M$ whose tangent bundle $TM$ admits a reduction of its structure group from ${\rm GL}(m,\R)$ to $H$.  Manifolds of dimension $2n$ with SU($n$)-structure admit a set of canonical
objects, consisting of an almost complex structure $J$, a Riemannian metric
$g$, a real two-form $\omega$ and a complex $n$-form $\Omega$.
With respect to $J$, the forms $\omega$ and $\Omega$ are of type (1,1)
and ($n$,0), respectively, and there is a compatibility condition,
$g(J\cdot,\cdot)=\omega(\cdot,\cdot)$.  With respect to the volume
form $V_g$ of $g$, $\omega$ and $\Omega$ are normalised so that
\begin{equation}
\omega^n \= n!V_g \und
\Omega\wedge\bar{\Omega} \= (2\im)^n V_g\ .
\end{equation}
Let $e^a$, where $a=1,\dots,2n$, be a local frame for the cotangent bundle,
and let $\Theta^\mu=e^{2\mu-1}+\im e^{2\mu}$ and
$\Theta^{\bar\mu}:=\bar{\Th}^\mu$ be local frames for the (1,0) and
(0,1) parts of the complexified cotangent bundle. Then $J,\ome,\Om$ may be written as follows:
\begin{equation}
\label{SU3 structure}
J\Th^\mu=\im\Th^\mu\ , \quad \ome=
\frac{\im}{2}\sum_{\mu=1}^n\Th^\mu\wedge\Th^{\bar\mu}\ ,
 \quad \Om=\Th^1\wedge\Th^2\wedge\dots\wedge\Th^n\ .
\end{equation}

One may choose a connection which preserves $g$, $J$, $\ome$ and $\Om$.
We are particularly interested in the case where the torsion of this
connection is totally anti-symmetric, so that the Cartan structure
equations may be written
\begin{equation}
\label{Cartan}
\diff\Th^\mu = \Gamma^\mu_\nu\wedge\Th^\nu + \Th^\mu\lrc T\ .
\end{equation}
Here $\Gamma$ encodes the Christoffel symbols of the connection: it is a
traceless anti-Hermitian matrix with 1-form entries.  The 3-form $T$ is
the torsion of the connection, while the symbol $\lrc$ denotes contraction,
defined in terms of the Hodge star by $u\lrc v = \ast(u\wedge\ast v)$.

Two particularly interesting cases arise when the torsion $T$ is the real
part of either a (3,0)-form or a (2,1)-form.  In the former case, the
manifold is called nearly K\"ahler \cite{WG,Gray,But}.  The latter case is equivalent
to a property called Calabi-Yau torsion \cite{CYT}, or CYT, as we now explain.

On a Hermitian manifold, the K\"ahler-torsion (or Bismut) connection is a
unitary connection $\tilde\Gamma$ on the cotangent bundle with anti-symmetric
torsion given by $\tilde T=J\diff\omega$ (where $J$ acts on the 3-forms
according to the rule $J(u\wedge v\wedge w) = (Ju\wedge Jv \wedge Jw)$).
A Hermitian manifold is called CYT when the K\"ahler-torsion (KT) connection
has structure group SU($n$) rather than U($n$) (so that $\tilde\Gamma$ is
traceless as well as anti-Hermitian).

Now suppose that the torsion $T$ of an SU($n$)-structure on some
$2n$-manifold is the real part of a (2,1)-form.  The right hand side of
(\ref{Cartan}) contains no part of type (0,2).  This immediately tells
us that the almost complex structure $J$ is integrable, and the manifold
is Hermitian.  Writing
\begin{equation}
\label{cyt torsion 1}
T=\sfrac{1}{2}
T_{\mu\nu\bar{\r}}\Theta^\mu\wedge\Theta^\nu\wedge\Th^{\bar\r}+
\sfrac{1}{2}
T_{\mu\bar{\nu}\bar{\r}}\Theta^\mu\wedge\Th^{\bar\nu}\wedge\Th^{\bar\r}\ ,
\end{equation}
and using the Cartan structure equation (\ref{Cartan}) and the
anti-Hermiticity of $\Gamma$, one can show that
\begin{eqnarray}
\label{cyt torsion 2}
\diff\omega
&=& \sfrac{\im}{2} \left( (\Gamma^\mu_\nu\wedge\Th^\nu+
\Th^\mu\lrc T)\wedge\Th^{\bar\mu} -
\Th^\mu\wedge(\Gamma^{\bar\mu}_{\bar\nu}\wedge\Th^{\bar\nu}+
\Th^{\bar\mu}\lrc T) \right) \\
&=& -\sfrac{\im}{2}T_{\mu\nu\bar{\r}}\Theta^\mu\wedge\Theta^\nu\wedge\Th^{\bar\r}+
\sfrac{\im}{2}T_{\mu\bar{\nu}\bar{\r}}\Theta^\mu\wedge\Th^{\bar\nu}\wedge\Th^{\bar\r} \\
&=& -JT\ .
\end{eqnarray}
Thus the K\"ahler-torsion connection $\tilde\Gamma$ and the connection
$\Gamma$ preserving the SU($n$)-structure have the same torsion
$T=\tilde T=J\diff\ome$.  So these two connections must agree, and
the manifold is CYT.

\subsection{Intrinsic torsion classes in dimension six}

Now we restrict attention to the case of 6-manifolds with SU(3)-structure.
Chiossi and Salamon have provided a classification of SU(3)-structures in
terms of intrinsic torsion classes \cite{Chiossi:2002tw}.  Here we explain
how nearly K\"ahler manifolds and CYT manifolds fit into this classification.  The five torsion classes can be determined by analysing either the torsion tensor or the exterior derivatives of $\ome$ and $\Om$.  Here we follow the latter approach: the torsion classes are defined by
\begin{eqnarray}
\label{torsion classes 1}
\diff\omega &=& {\rm Im}((W_1^+-\im W_1^-)\Om) + W_3 + W_4\wedge\omega \ ,\\
\label{torsion classes 2}
\diff\Om &=& (W_1^++\im W_1^-)\omega\wedge\omega + (W_2^++\im W_2^-)\wedge\omega + \Om\wedge W_5\ .
\end{eqnarray}
Here $W_1^\pm$ are real functions, $W_4,W_5$ are real 1-forms, $W_2^\pm$ are the real and imaginary parts of a (1,1)-form with $\omega\lrc W_2^\pm=0$, and $W_3$ is the real part of a (2,1)-form with $\omega\lrc W_3=0$.

The torsion classes of a nearly K\"ahler manifold are already well-known.
Since the space of (3,0)-forms is 1-dimensional, one can write
$T={\rm Re}\,\rho\,\Om$ for some complex function $\rho$.
Using (\ref{Cartan}) one can calculate
\begin{eqnarray}
\diff \ome &=& \sfrac{\im}{2}
\left( ( \Gamma^\a_\b\wedge\Th^\b + \Th^\a\lrc {\rm Re}\,\rho\,\Om)\wedge\Th^{\bar\a}
- \Th^\a\wedge (\Gamma^{\bar\a}_{\bar\b}\wedge\Th^{\bar\b} +
\Th^{\bar\a}\lrc {\rm Re}\,\rho\,\Om)\right ) \\
&=& 3\,{\rm Im}\,\rho\,\Om \ , \\
\diff \Om &=& \sfrac{1}{2}\eps_{\a\b\g} (\Gamma^\a_\de\wedge \Th^\de + \Th^\a\lrc {\rm Re}\,\rho\,\Om) \wedge\Th^\b\wedge\Th^\g \\
&=& 2\bar\rho\,\ome\wedge\ome \ .
\end{eqnarray}
Comparing with (\ref{torsion classes 1}), (\ref{torsion classes 2}), one sees that 
\begin{equation} W_2^\pm=W_3=W_4=W_5=0 \end{equation}
for a nearly K\"ahler 6-manifold.

The torsion classes for CYT manifolds can be derived similarly (see also \cite{Lopes Cardoso:2002hd}).  Since the complex structure $J$ is integrable, $\diff\Omega$ is a (3,1)-form and the torsion classes $W_1$ and $W_2$ vanish.  Let $\tilde{\Gamma}$ denote the KT connection, with torsion $\tilde{T}$.  Then $\tilde{\Gamma}$ is a Hermitian matrix with 1-form entries and $\tilde{\Gamma},\tilde T$ satisfy the Cartan structure equations (\ref{Cartan}).  The manifold will be CYT if and only if $\tilde\Gamma$ is traceless.  One may calculate $\diff\Om$ as follows:
\begin{eqnarray}
\diff\Omega &=&
\sfrac{1}{2}\epsilon_{\a\b\g}(\tilde\Gamma^\a_\de\wedge\Theta^\de+
\Theta^\a \lrc \tilde T)\wedge\Theta^\b\wedge\Theta^\g \\
&=& \delta^\de_\a(\tilde\Gamma^\a_\de +
2\de^{\a\bar\a}\tilde T_{\bar{\a}\bar{\b}\de}{\Theta}^{\bar\b})\wedge\Omega \\
&=& (\tilde\Gamma^\a_\a + \omega\lrc\diff\omega)\wedge\Omega\ .
\end{eqnarray}
We recall that $W_4$ and $W_5$ are determined by $W_4=\sfrac{1}{2}\omega\lrc\diff\omega$ and $\diff\Omega = \Omega\wedge W_5$.  So $\tilde\Gamma$ is traceless if and only if $2W_4+W_5=0$.  Therefore a CYT manifold satisfies
\begin{equation}
W_1=0\ ,\quad W_2^\pm=0\ ,\quad 2W_4+W_5=0\ .
\end{equation}
The CYT property does not place any restriction on $W_3$.

\section{The Spin(7)-instanton equations}
\label{sec:3}

\subsection{Spin(7)-structure and instanton equations}

A Spin(7)-structure on an 8-manifold is specified by a self-dual 4-form
$\Sigma$ satisfying certain conditions.  Given any SU(4)-structure on an
8-manifold, there is an $S^1$'s-worth of compatible Spin(7)-structures,
determined by
\begin{equation}
\label{Spin(7)}
\Sigma = \mathrm{Re}\, e^{\im\xi}\Om + \sfrac{1}{2} \omega\wedge\omega\ ,
\end{equation}
with $\xi$ a real parameter.  Conversely, given such a Spin(7)-structure,
the space of compatible SU(4)-structures is ${\rm Spin}(7)/{\rm SU}(4)=
{\rm Spin}(7)/{\rm Spin(6)}=S^6$.

A natural generalisation of the 4-dimensional anti-self-dual Yang-Mills equation to an
8-dimensional manifold with Spin(7)-structure is the $\Sigma$-anti-self-dual equation:
\begin{equation}
\label{SigmaASD}
\ast \Fcal = -\Sigma\wedge\Fcal\ .
\end{equation}
When the Spin(7)-structure is determined by an SU(4)-structure via
(\ref{Spin(7)}), the equations can be written \cite{DT,Tian}:
\begin{eqnarray}
\label{SU4inst1}
e^{-\im\xi}\mathcal{F}_{\mu\nu} &=& -\sfrac{1}{2} \epsilon_{\mu\nu\rho\sigma}
 \mathcal{F}_{\bar{\rho}\bar{\sigma}}\ , \\
\label{SU4inst2}
\omega \lrc \mathcal{F} &=& 0\ .
\end{eqnarray}
Here we have written the field strength tensor $\Fcal$ in a basis of (1,0)- and (0,1)-forms:
\begin{equation}
\Fcal = \sfrac{1}{2} \Fcal_{\mu\nu}\Th^\mu\wedge\Th^\nu +
\Fcal_{\mu\bar\nu}\Th^\mu\wedge\Th^{\bar\nu} +
\sfrac{1}{2}\Fcal_{\bar\mu\bar\nu}\Th^{\bar\mu}\wedge\Th^{\bar\nu}\ .
\end{equation}
Anti-hermiticity of $\Fcal$ implies that $\Fcal_{\bar\mu\bar\nu}=
-\Fcal_{\mu\nu}^\dagger$ and $\Fcal_{\mu\bar\nu}^\dagger=\Fcal_{\nu\bar\mu}$.

Note that the Spin(7)-instanton equations are weaker than the
Hermitian-Yang-Mills equations: 
\begin{eqnarray}
\Fcal_{\mu\nu}&=&0\ ,\\
\omega\lrc\Fcal&=&0\ .
\end{eqnarray}
Any solution of the Hermitian-Yang-Mills equations is automatically a
solution of the Spin(7)-instanton equations.

We will sometimes need to be able to determine whether two SU(4)-structures are compatible with the same fixed Spin(7)-structure.  The (1,1)-form $\tilde\omega$ of a compatible SU(4)-structure should satisfy
\begin{equation}
\label{Sigma SD}
\ast\tilde\omega = \sfrac{1}{3}\Sigma\wedge\tilde\omega.
\end{equation}
The space of solutions to this equation is 7-dimensional, so the space of normalised solutions is $S^6$.  Any solution $\tilde\ome$ of (\ref{Sigma SD}) is a linear sum of $\omega$ and a solution $f$ of the complex self-dual equation:
\begin{equation}
\label{complex SD}
e^{-\im\xi}f_{\mu\nu} = \sfrac{1}{2} \epsilon_{\mu\nu\rho\sigma}
 f_{\bar{\rho}\bar{\sigma}}\ .
\end{equation}

\subsection{Yang-Mills action functional}

Taking the exterior derivative of the $\Sigma$-anti-self-dual equation (\ref{SigmaASD}) and using
the Bianchi identity gives
\begin{equation}
\label{YMEtorsion}
D\ast \Fcal + \Fcal\wedge\diff\Sigma = 0\ ,
\end{equation}
where $D$ is the covariant derivative and $\diff\Sigma$ is Hodge dual to a torsion 3-form.  The torsionful Yang-Mills equation (\ref{YMEtorsion}) is the variational equation for the action,
\begin{equation}
\label{YMAtorsion}
S=-\int {\rm Tr}\left(\Fcal\wedge\ast\Fcal +
\Fcal\wedge\Fcal\wedge\Sigma\right)\ .
\end{equation}
The $\Sigma$-anti-self-dual equation (\ref{SigmaASD}) can be derived from this action using a Bogomolny argument: there is a lower bound on the action,
\begin{equation}
\label{Bogomolny bound}
S\geq 0\ ,
\end{equation}
which is saturated if and only if (\ref{SigmaASD}) holds.

The proof of (\ref{Bogomolny bound}) in the integrable case $\diff\Sigma=0$ appeared in \cite{Bau}, and is easily adapted to the case $\diff\Sigma\neq0$.  One can orthogonally decompose the curvature $\Fcal=\Fcal^++\Fcal^-$ into two parts satisfying $\Sigma\wedge \Fcal^-=-\ast \Fcal^-$ and $\Sigma\wedge \Fcal^+=3\ast \Fcal^+$.  The lower bound follows from integrating the identity,
\begin{eqnarray}
-{\rm Tr}(\Fcal\wedge\ast\Fcal+\Fcal\wedge\Fcal\wedge\Sigma) = - 4\,{\rm Tr}(\Fcal^+\wedge\ast\Fcal^+)\ ,
\end{eqnarray}
and discarding the positive piece $-\int{\rm Tr}(\Fcal^+\wedge\ast\Fcal^+)$.  Clearly the inequality is satisfied if and only if $\Fcal^+=0$.

\subsection{Reduction to four dimensions and the geometric Langlands program}
\label{sec:KW}

Now we will consider some translation-invariant cases of the Spin(7)-instanton equations on $\R^8$.

Let $x^\mu$ and $y^a$ be coordinates on $\R^n$ and $\R^{8-n}$, with $\mu=1,\dots,n$ and $a=1,\dots,8-n$.  A gauge field invariant under $y$-translations is written,
\begin{equation}
\label{flat gauge field}
\Acal = A + \Phi = A_\mu(x)\diff x^\mu + \phi_a(x)\diff y^a\ .
\end{equation}
Its curvature is
\begin{equation}
\label{flat field strength}
\Fcal = F + D\Phi + \Phi\wedge\Phi\ ,
\end{equation}
with $F$ denoting the curvature of $A$ and $D\Phi = (\diff\phi_a+[A,\phi_a])\diff y^a$ a covariant derivative.

Consider the following SU(4)-structure in the case $n=4$:
\begin{equation}
\Th^\mu = \diff x^\mu + \im \diff y^\mu\ .
\end{equation}
The (2,0)-part of the field strength is
\begin{equation}
\label{KW0}
\Fcal^{2,0} = \frac{1}{8}\left( F_{\mu\nu}-[\phi_\mu,\phi_\nu] - \im(D_\mu\phi_\nu-D_\nu\phi_\mu) \right) \Th^\mu\wedge\Th^\nu.
\end{equation}
If we write $\chi=\phi_\mu\diff x^\mu$, the complex anti-self-dual
equation (\ref{SU4inst1}) can be cast neatly as
\begin{eqnarray}
\label{KW1}
\left(F-\chi\wedge\chi \right) +
t D \chi && \mbox{ is anti-self-dual}\ , \\
\label{KW2}
D \chi - t\left(F-\chi\wedge\chi
\right) && \mbox{ is self-dual}\ ,
\end{eqnarray}
where $t=-\tan(\xi/2)$.  The second of the Spin(7)-instanton equations (\ref{SU4inst2}) is
\begin{equation}
\label{KW3}
D \ast \chi = 0\ .
\end{equation}
The Hermitian-Yang-Mills equations are slightly stronger: they consist of (\ref{KW3}), and the vanishing of the (2,0)-part (\ref{KW0}) of $\Fcal$:
\begin{eqnarray}
\label{KW4}
F-\chi\wedge\chi &=&0\ , \\
\label{KW5}
D \chi&=&0\ .
\end{eqnarray}

Equations (\ref{KW1})-(\ref{KW5}), generalised to a curved manifold, were used by Kapustin and Witten in their work on the geometric Langlands program \cite{Kapustin:2006pk}.  The Spin(7)-instanton equations (\ref{KW1})-(\ref{KW3}), with complex parameter $t$, define a topological quantum field theory.  When $t$ is real these are just the Spin(7)-instanton equations; when $t$ is non-real they are equivalent, by taking Hermitian and anti-Hermitian parts, to the Hermitian-Yang-Mills equations (\ref{KW3})-(\ref{KW5}).  So all of Kapustin and Witten's equations have a natural interpretation as 8-dimensional anti-self-dual equations.

Equations (\ref{KW1})-(\ref{KW3}) have also recently found an application in an analytically continued Chern-Simons theory.  We will have more to say about this at the end of this section.

\subsection{Reduction to two dimensions and the Hitchin and Nahm equations}

Now we consider the case $n=2$.  Let $z=x^1+\im x^2$ be the complex coordinate on $\R^2=\C$ and let $w^\a=y^{2\a-1}+\im y^{2\a}$, $\a=1,2,3$, be the complex coordinates on $\C^3$.  The natural SU(4)-structure (with
vanishing torsion) is
\begin{equation}
\Theta^\a=\diff w^\a,\ \a=1,2,3\ , \quad \Theta^4=\diff z\ .
\end{equation}
Throughout this paper we will adopt a convention that indices $\a,\b,\g$ run
from 1 to 3.  We again impose translational symmetry via equation (\ref{flat gauge field}), with 
\begin{eqnarray}
\Phi = \phi_\a(z,\bar z) \Th^\a + \phi_{\bar \a}(z,\bar z) \Th^{\bar\a}\, .
\end{eqnarray}
So $A$ is a gauge field on $\C$ and $\phi_1,\phi_2,\phi_3$ are matrix-valued functions, with $\phi_{\bar \a}=-\phi_\a^\dagger$ for anti-Hermiticity.

The Spin(7)-instanton equations (\ref{SU4inst1}), (\ref{SU4inst2}) for $\Fcal$ reduce to
\begin{eqnarray}
\label{gen hitchin 1}
e^{\im\xi}D_{\bar{z}}\phi_{\bar{\a}} &=&
\sfrac{1}{2}\epsilon_{\bar\a\bar\b\bar\g}[\phi_\b,\phi_\g]\ , \\
\label{gen hitchin 2}
\ast F &=& 2\im [\phi_\a,\phi_{\bar{\a}}]\ .
\end{eqnarray}
These equations generalise both the Hitchin equations and the Nahm equations.  The Hitchin equations are recovered on setting $\phi_2=\phi_3=0$:
\begin{eqnarray}
D_{\bar{z}}\phi_{\bar{1}} &=& 0\ , \\
\ast F &=& 2\im [\phi_1,\phi_{\bar{1}}]\ .
\end{eqnarray}
The Nahm equations arise when $\phi_\a$ are anti-Hermitian: in this case equation (\ref{gen hitchin 2}) implies that $F=0$, so that $A=0$ in some gauge.  Then equation (\ref{gen hitchin 1}) (with $\xi=0$) implies that $\pa_2\phi_\a=0$ and
\begin{equation}
\pa_1\phi_{\a} = \epsilon_{\a\b\g}[\phi_\b,\phi_\g].
\end{equation}

\subsection{Hermitian flow equations}

One further interesting property of equation (\ref{gen hitchin 1}) is that it is a ``Hermitian flow'' on a Hermitian manifold.  A Hermitian flow is the analogue for a Hermitian manifold of gradient flow on a Riemannian manifold or Hamiltonian flow on a symplectic manifold.  Suppose that we have a Hermitian manifold $N$ with coordinates $u^\mu$ and Hermitian metric $h_{\mu\bar\nu}\diff u^\mu\diff u^{\bar\nu}$.  Given a function $W$ on $N$, one can define a Hermitian flow equation
for maps from $\C$ into $N$:
\begin{equation}
\frac{\partial u^{\bar\mu}}{\partial\bar z} =
h^{\bar\mu\nu}\frac{\partial W}{\partial u^\nu}\ .
\end{equation}
The flow equation can be cast in a coordinate-free fashion as follows:
\begin{equation}
h\left( X,\frac{\partial}{\partial\bar z}\right) =
\partial W(X) \qquad\forall X\in \Gamma(T^{1,0}N)\ .
\end{equation}

Consider the space $N$ of triples of complex matrices $(\phi_1,\phi_2,\phi_3)$.  Holomorphic and anti-holomorphic tangent vectors are written 
\begin{equation}
X=X_\a\frac{\pa}{\pa\phi_\a}\ ,\quad \bar X=X_{\bar\a}\frac{\pa}{\pa\phi_{\bar\a}}\ ,
\end{equation}
with $(X_1,X_2,X_3)$ again a triple of complex matrices and $X_{\bar\a}=-X_\a^\dagger$.  A natural Hermitian metric is
\begin{equation}
h(X,\bar Y) = -{\rm Tr}(X_\a Y_{\bar\a}).
\end{equation}
Then the Hermitian flow for the real function
\begin{equation}
W(\phi) = -\sfrac{1}{3}{\rm Tr}(\eps^{\a\b\g}\phi_\a\phi_\b\phi_\g + \eps^{\bar\a\bar\b\bar\g}\phi_{\bar\a}\phi_{\bar\b}\phi_{\bar\g})
\end{equation}
is
\begin{equation}
\frac{\pa\phi_{\bar\a}}{\pa\bar{z}} =
\sfrac{1}{2}\epsilon_{\bar\a\bar\b\bar\g}[\phi_\b,\phi_\g]\ ,
\end{equation}
and this coincides with (\ref{gen hitchin 1}) when $A=0$.

The fact that (\ref{gen hitchin 1}) is a Hermitian flow reflects the more general property that the complex anti-self-dual equation (\ref{SU4inst1}) on $\C\times M^6$ is, for suitable $M^6$, a Hermitian flow on the infinite-dimensional space of connections on $M^6$.  The space $N$ of connections on
$M^6$ inherits a natural complex structure from that on $M^6$, and
holomorphic and anti-holomorphic vector fields on $N$ are (1,0)- and (0,1)-forms
on $M^6$, respectively.  A natural Hermitian metric is given by
\begin{equation}
h(\psi,\eta) = -4\int_{M^6}{\rm Tr}(\psi\wedge\ast\eta) \for
\psi\in\Lambda^{1,0}M^6\ ,\quad \eta\in\Lambda^{0,1}M^6 \ .
\end{equation}
If ${\rm Im}\Om$ is closed, then the Hermitian flow equation for the function
\begin{equation}
 W(\Acal) = -\int_{M^6} {\rm Tr}\left(\Acal\wedge \diff \Acal +
 \sfrac{2}{3}\Acal\wedge \Acal\wedge \Acal\right) \wedge{\rm Im}\Om
\end{equation}
is
\begin{equation}
\frac{\partial \Acal_{\bar\a}}{\partial\bar z} =
\frac{1}{2}\,\eps_{\bar\a\bar\b\bar\g}\Fcal_{\b\g}\ .
\end{equation}
This equation is equivalent to the complex anti-self-dual equation (\ref{SU4inst1}) (with $\Acal_{\bar z}=0$).  The condition that ${\rm Im}\Om$ is closed implies that the torsion classes $W_3,W_5$ vanish: thus Calabi-Yau and nearly K\"ahler manifolds lead to a Hermitian flow, but CYT manifolds only do so if $W_5=0$.

We will close this section by making a few more comments on the 4-dimensional Kapustin-Witten equations (\ref{KW1})-(\ref{KW3}).  In a recent paper \cite{Witten:2010cx} these were found to play a fundamental role in an analytically continued Chern-Simons theory.  Witten observed that they emerge as gradient flow equations for the real part ${\rm Re}(\mathcal{I})$ of a holomorphic Chern-Simons functional $\mathcal{I}$ for a complex connection on a 3-manifold.  The above discussion explains why this should be the case: ${\rm Re}(\mathcal{I})$ arises by reducing $W$ from six to three dimensions.  The complex connection is $A+\im\phi$ in the language of subsection \ref{sec:KW}, and the Hermitian flow in two dimensions reduces to a gradient flow in one dimension.

\section{Coset spaces and invariant gauge fields}
\label{sec:4}

The standard example of a manifold with $H$-structure is a coset space $G/H$.  All of the manifolds that we will consider will be products of coset spaces $G/H$ with tori $T^d$ (or decompactified tori $T^{d-p}\times \R^p$), such that the subgroup $H$ of $G$ can be embedded in ${\rm SU}(4)$.  The natural $H$-structure can then be lifted to an SU(4)-structure, which in most of our examples is either nearly K\"ahler or CYT.

In the first part of this section we will describe the natural $H$-structure on a coset space $G/H$, and in the second part we will describe the machinery required to write down $G$-invariant gauge fields on $G/H$.  Standard references for this material are \cite{Kobayashi-Nomizu1,KVMR,Zoupanos}.

\subsection{Coset space geometry}

Let $G/H$ denote the space of left cosets $gH$.  We choose a frame for the
cotangent bundle of $G/H$ as follows.  Assume that there exists a subspace
$\mathfrak{m}$ of $\mathfrak{g}$ such that $\mathfrak{g}=
\mathfrak{h}\oplus\mathfrak{m}$ and $[\mathfrak{h},\mathfrak{m}]\subset
\mathfrak{m}$, where $\mathfrak{g},\mathfrak{h}$ denote the Lie algebras of $G$
and $H$.  Let $I_a$ and $I_i$ be bases for $\mathfrak{m}$ and $\mathfrak{h}$
respectively, where $a=1,\dots,\dim G/H$ and $i=\dim G/H+1,\dots,\dim G$.
These will be chosen orthonormal with respect to an adjoint-invariant quadratic form $q$, for example a multiple of the Cartan-Killing form.

The basis elements $I_a,I_i$ determine left-invariant vector fields on $G$,
whose dual left-invariant 1-forms are denoted $\hat e^a,\hat e^i$.
Over a coordinate patch $U\subset G/H$, we can choose a local section $L$ of
the principle bundle $G\rightarrow G/H$ (in other words, a map
$L:U\rightarrow G$ such that $\pi\circ L$ is the identity, where
$\pi:G\rightarrow G/H$ is the natural projection).  The 1-forms $\hat e^a,
\hat e^i$ can be pulled back to 1-forms $e^a,e^i$ on $U$ by the local section.

The 1-forms $e^a$ form a local frame for the cotangent bundle of $G/H$.  We can define a metric on $G/H$ by declaring them to be orthonormal: this is the natural metric induced by the chosen quadratic form $q$ on $\mathfrak{g}$.  The remaining 1-forms $e^i$ define the so-called canonical connection,
\begin{equation}
\label{canonical connection}
 A^0=e^iI_i\ .
\end{equation}
This is a $G$-invariant connection on the $H$-bundle $G\rightarrow G/H$; there is a one-to-one correspondence between $G$-invariant connections and choices of $\mathfrak{m}\subset\mathfrak{g}$.  Since $e^a$ is a frame, one can write $e^i=e^i_ae^a$.  The differentials of the 1-forms $e^i,e^a$ are encoded in the Maurer-Cartan equations:
\begin{eqnarray}
\label{MC1}
\diff e^a &=& -f_{ib}^a\;e^i\wedge e^b\ -\ \sfrac12 f_{bc}^a\;e^b\wedge e^c\ , \\
\label{MC2}
\diff e^i &=&-\sfrac12 f_{bc}^i\,e^b\wedge e^c\ -\ \sfrac12 f_{jk}^i\,e^j\wedge e^k\ .
\end{eqnarray}
Here the $f$'s are the structure constants of the Lie algebra $\mathfrak{g}$, defined by
\begin{equation}
 [I_i,I_j]=f_{ij}^k I_k\ ,\quad
 [I_i,I_a]=f_{ia}^bI_b\ ,\quad
 [I_a,I_b]=f_{ab}^iI_i + f_{ab}^c I_c\ .
\end{equation}

The first Maurer-Cartan equation (\ref{MC1}) can be interpreted as a Cartan structure
equation (\ref{Cartan}), written in real form:
\begin{equation}
\diff e^a = - \Gamma^a_b\wedge e^b + e^a\lrc T\ .
\end{equation}
The connection $\Gamma^a_b=e^if^a_{ib}$ is just the canonical connection (\ref{canonical connection}) acting on 1-forms via the adjoint action of $\mathfrak{h}$ on $\mathfrak{m}$, so its structure group is $H$.  Its torsion
\begin{equation}
\label{torsion}
T=-\sfrac16\,f_{abc}\, e^a\wedge e^b\wedge e^c
\end{equation}
is a 3-form, because the structure constants $f_{abc}=\delta_{ad}f^d_{bc}$ are
totally antisymmetric in $a,b,c$.

Because the connection $\Gamma$ has structure group $H$, there is an underlying $H$-structure on $G/H$.  Globally this comes about because the cotangent bundle $T^\ast(G/H)$ is isomorphic to the vector bundle $G\times_H\mathfrak{m}^\ast$ associated to the $H$-principle bundle $G\rightarrow G/H$ via the adjoint action of $H$ on the dual $\mathfrak{m}^\ast$ of $\mathfrak{m}$.

\subsection{$G$-invariant gauge fields}

Bundles over $G/H$ to which the action of $G$ can be lifted are determined
by homomorphisms from $H$ into the gauge group $K$.  The images of the
generators $I_i$ of $H$ will be denoted $L_i$, while the whole set of
generators of $K$ will be denoted $L_A$ and the structure constants will
be denoted $h_{AB}^C$.  Gauge fields invariant under the action of $G$ are
determined by linear maps from $\mathfrak{m}$ to the Lie algebra $\mathfrak{k}$ of $K$, which commute with the action of $H$.  Such maps can be written
$I_a\mapsto \Phi_a^AL_A$, and $\Phi_a^A$ must satisfy
\begin{equation}
\label{Phi}
f_{ia}^b \Phi_b^A = h_{iB}^A\Phi_a^B\ .
\end{equation}

The most general possible $G$-invariant gauge field on $M\times G/H$ can
be written
\begin{equation}
\label{coset gauge field}
\Acal = A + A^0 + L_A\Phi^A_a e^a \ .
\end{equation}
Here $A$ denotes a gauge field on the smooth manifold $M$ of dimension $d$, with gauge group equal to the
centraliser of the image of $H$ in $K$.  The canonical connection is denoted
$A^0=e^i L_i$.  We also use $\Phi$ as shorthand for $L_A\Phi^A_a e^a$, where
$\Phi^A_a$ satisfies (\ref{Phi}) and is allowed to depend on the coordinate
on $M$.

In order to calculate the field strength tensor, we first split the exterior
derivative into two parts acting on $M$ and $G/H$:
$\diff = \diff^{M}+\diff^{G/H}$.  Then short calculations using the
Maurer-Cartan equations (\ref{MC1}), (\ref{MC2}) and the identity (\ref{Phi})
show that
\begin{eqnarray}
F^0:= \diff^{G/H} A^0 + A^0\wedge A^0 &=&
-\sfrac12 f_{ab}^i L_i e^a\wedge e^b\ , \\
\diff^{G/H} \Phi + A^0\wedge \Phi + \Phi\wedge A^0 &=& \Phi \lrc T\ ,
\end{eqnarray}
with $T$ the torsion tensor (\ref{torsion}).  Thus the field strength
of $\Acal$ is given by
\begin{equation}
\label{coset field strength}
\Fcal = F + F^0 + D(\Phi^A_aL_A) \wedge e^a +\Phi\wedge\Phi +\Phi\lrc T\ .
\end{equation}
Here we have introduced the covariant derivative $D=d^{M} + [A,\cdot]$.

The field strength (\ref{coset field strength}) should be compared with the translation-invariant field strength on flat space (\ref{flat field strength}).  In fact (\ref{coset field strength}) differs
from (\ref{flat field strength}) in precisely three ways: by the field
strength $F^0$ of the canonical connection, by the torsion term involving $T$,
and by the constraint (\ref{Phi}) imposed on $\Phi$.  In most of the cases
that we consider, $F^0$ solves the Spin(7)-instanton equations, so that the
Spin(7)-instanton equations will differ from the relevant flat cases only by the torsion term and the constraint on $\Phi$.

For the remainder of this section, we will explain why $F^0$ solves
the Spin(7)-instanton equations whenever the subgroup $H$ of $G$ can be embedded in SU(4).  Actually, we will show that it solves the stronger Hermitian-Yang-Mills equations.  A more explicit proof for the nearly K\"ahler case can be found in \cite{HILP}.

One way to describe the
Hermitian-Yang-Mills equations for a 2-form $F$ in 8 dimensions is to say that $F$ belongs to the Lie
sub-algebra $\mathfrak{su}(4)$, when 2-forms are identified with elements of
$\mathfrak{so}(8)$ via the metric.  We wish to show that the 2-forms
$(1/2)f_{ab}^ie^a\wedge e^b$ solve the Hermitian-Yang-Mills equations.
These map to matrices $B_i\in\mathfrak{so}(8)$ with entries $f_{ia}^b$.
Then it is obvious that these $B_i$ belong to the sub-algebra
$\mathfrak{su}(4)$: the $B_i$ are the generators of
$\mathfrak{h}\subset\mathfrak{su}(4)$, acting on vectors via the adjoint
action of $\mathfrak{h}$ on $\mathfrak{m}$.

Note that, if we choose a metric on $G/H$ other than that induced by the Cartan-Killing form, $F^0$ is no longer guaranteed to solve the Spin(7)-instanton equations.  We will see examples of metrics for which $F^0$ does not solve the Spin(7)-instanton equations in section \ref{sec:8}.

\section{Yang-Mills fields on parallelisable CYT spaces}
\label{sec:5}

\subsection{The Calabi-Eckmann complex structure}

Our basic examples of CYT manifolds will be products of two odd-dimensional spheres, which may be regarded as a coset spaces:
\begin{equation}
S^{2n-1}\times S^{2m-1} = {\rm S}({\rm U}(n)\times{\rm U}(m)) / {\rm S}({\rm U}(n-1)\times{\rm U}(m-1))\, \quad\mbox{with } n,m\geq1 \ .
\end{equation}
These admit natural ${\rm S}({\rm U}(n-1)\times{\rm U}(m-1))$-structures, which may be lifted to CYT ${\rm SU}(n+m-1)$-structures.  There is actually a choice of lift; in our examples we will consider 2-parameter families of CYT structures.

The Calabi-Eckmann complex structure can loosely be described as follows: the spheres $S^{2n-1}$, $S^{2m-1}$ are regarded as $S^1$-fibrations over $\mathbb{CP}^{n-1}$, $\mathbb{CP}^{m-1}$.  The projective spaces are each equipped with their standard complex structure, while a suitable complex structure is chosen on the fibre $S^1\times S^1$.  Thus these CYT spaces are elliptic fibrations, so are perhaps suitable compactification spaces for F-theory.

The natural metric to choose on $S^{2n-1}={\rm SU}(n)/{\rm SU}(n-1)$ is induced by the Cartan-Killing form on $\mathfrak{su}(n)$, when $n\geq2$ ($S^1$ is equipped with the usual round metric).  It is worth emphasising that this metric coincides with the usual round metric only in the cases of $S^3$ and $S^1$.  We will consider 2-parameter families of product metrics on $S^{2n-1}\times S^{2m-1}$, obtained by rescaling the metrics on each factor.  Along with a family of compatible complex structures, these will determine families of CYT structures.

In this section we will consider the Spin(7)-instanton equations on 2 examples of 8-dimensional CYT manifolds whose natural structure group is the trivial group.  They are $S^3\times T^5$ and $S^3\times S^3\times T^2$.  In the second case there is more than one way to obtain a CYT structure: one may regard it either as a product of two copies of CYT $S^3\times S^1$, or as a product of $S^3\times S^3$ with a torus.  All gauge and scalar fields will take values in the Lie algebra $\mathfrak{k}$ of the gauge group $K$, or its complexification $\mathfrak{k}_\C$.

\subsection{Invariant gauge fields on $(S^3\times S^3)\times T^2$}

The manifold $S^3\times S^3$ is the quotient of ${\rm SU}(2)^2$ by the trivial subgroup $H=\{1\}$, so fits into the discussion of section \ref{sec:4}, albeit as a slightly trivial case.

Let $\hat e^a,\check e^a,a=1,2,3$, be left-invariant 1-forms on two copies of ${\rm SU}(2)=S^3$, normalised so that $\diff\hat e^a=-(1/2R_1)\eps_{abc}\hat e^b\wedge\hat e^c$ and $\diff\check e^a=-(1/2R_2)\eps_{abc}\check e^b\wedge\check e^c$, with $R_1,R_2$ positive real constants which will determine the radii of the two 3-spheres.  Let $z$ denote a complex coordinate on $T^2$.  An SU(4)-structure on $S^3\times S^3\times T^2$ is defined by setting
\begin{equation}
\Th^1=\hat e^1+\im \hat e^2\ , \quad
\Th^2=\check e^1+\im\check e^2\ , \quad
\Th^3=\hat e^3+\im\check e^3\ , \quad \Th^4 = \diff z.
\end{equation}
The SU(4)-structure is preserved by the canonical connection, whose Christoffel symbols $\Gamma$ vanish and whose torsion (\ref{torsion}) is
\begin{eqnarray}
\label{S3xS3 torsion}
T &=& -\frac{1}{R_1}\hat e^1\wedge\hat e^2\wedge\hat e^3 -\frac{1}{R_2} \check e^1\wedge \check e^2 \wedge \check e^3 \\
&=& -\frac{\im}{4R_1}\,\Th^1\wedge\Th^{\bar1}\wedge(\Th^3+\Th^{\bar3}) +
\frac{1}{4R_2}\,\Th^2\wedge\Th^{\bar2}\wedge(\Th^{\bar3}-\Th^3).
\end{eqnarray}
Since the torsion is the real part of a (2,1)-form, this SU(4)-structure is CYT.

We make the usual ansatz (\ref{coset gauge field}) for an ${\rm SU}(2)^2$-invariant gauge field $\Acal$, with $A^0=0$ and $A$ a gauge field on $T^2$ taking values in $\mathfrak{k}$.  We parametrise $\Phi$ in terms of scalars $\phi_\a$, $\a=1,2,3$, taking values in $\mathfrak{k}_\C$:
\begin{equation}
\Phi = \phi_\a \Th^\a + \phi_{\bar \a}\Th^{\bar\a}\ .
\end{equation}
Here and throughout this article we adopt the notation $\phi_{\bar\a}=-\phi_\a^\dagger$.  The field strength tensor $\Fcal$ of $\Acal$ is given in (\ref{coset field strength}) (with $T$ as in (\ref{S3xS3 torsion})).  The Spin(7)-instanton equations for $\Fcal$ on $S^3\times S^3\times T^2$ reduce to vortex-type equations,
\begin{eqnarray}
e^{\im\xi}D_{\bar z} \phi_{\bar1} &=&
[\phi_2,\phi_3] + \frac{1}{2R_2}\phi_2\ , \\
e^{\im\xi}D_{\bar z} \phi_{\bar2} &=&
[\phi_3,\phi_1] - \frac{\im}{2R_1}\phi_1\ , \\
e^{\im\xi}D_{\bar z} \phi_{\bar3} &=& [\phi_1,\phi_2]\ , \\
\ast F &=& 2\im[\phi_\a,\phi_{\bar\a}]+\left(\frac{1}{R_1}+\frac{\im}{R_2}\right)\phi_3+
\left(\frac{1}{R_1}-\frac{\im}{R_2}\right)\phi_{\bar 3} \ ,
\end{eqnarray}
on the complex torus $T^2$ (or one of its decompactifications $S^1\times\R$, $\R^2$).  These generalise equations (\ref{gen hitchin 1}), (\ref{gen hitchin 2}).  Here $D=d+[A,\cdot]$ denotes the covariant derivative on $T^2$.  Because $H$ is trivial, equation (\ref{Phi}) places no restrictions on $\phi_\a$.

\subsection{Invariant gauge fields on $(S^3\times S^1)^2$}
\label{sec:5.3}

An alternative choice of complex structure on $S^3\times S^3\times T^2$ is
\begin{equation}
\label{eq:5.3.0}
\Th^1 = \diff x^1+\im\hat e^3,\quad \Th^2 = \diff x^2+\im\check e^3,\quad \Th^3 = \hat e^1+\im\hat e^2,\quad \Th^4 = \check e^1+\im\check e^2\ ,
\end{equation}
with $x^1,x^2$ coordinates on the two copies of $S^1$.  This SU(4)-structure is also CYT, as can be seen from the canonical torsion:
\begin{eqnarray}
T &=& -\frac{1}{R_1}\hat e^1\wedge\hat e^2\wedge\hat e^3 -\frac{1}{R_2} \check e^1\wedge \check e^2 \wedge \check e^3 \\
&=& -\frac{1}{4R_1} \Th^3\wedge\Th^{\bar3}\wedge(\Th^1+\Th^{\bar1}) - \frac{1}{4R_2} \Th^4\wedge\Th^{\bar4}\wedge(\Th^2+\Th^{\bar2}) \ .
\end{eqnarray}
We parametrise the ${\rm SU}(2)^2$-invariant gauge field (\ref{coset gauge field}) as follows:
\begin{equation}
\Phi = \phi_i\Th^i + \phi_{\bar i}\Th^{\bar i} + \chi_1\hat e^3+\chi_2\check e^3, \quad i=3,4\ .
\end{equation}
Here $\chi_1,\chi_2$ take values in $\mathfrak{k}$, and $\phi_3,\phi_4$ take values in $\mathfrak{k}_\C$. Introducing $\chi=\chi_1\diff x^1+\chi_2\diff x^2$, the Spin(7)-instanton equations on $(S^3\times S^1)^2$ reduce to vortex-type equations,
\begin{eqnarray}
\label{eq:5.3.1}
\ast(F-\chi\wedge\chi-\im D\chi) &=& -4e^{\im\xi}[\phi_{\bar 3},\phi_{\bar 4}] \\
\label{eq:5.3.2}
D_1\phi_3 - \frac{\phi_3}{R_1} - \im[\chi_1,\phi_3] &=& e^{\im\xi}\left( D_2\phi_{\bar 4} - \frac{\phi_{\bar 4}}{R_2} + \im[\chi_2,\phi_{\bar 4}] \right) \\
\label{eq:5.3.3}
D_2\phi_3 - \im[\chi_2,\phi_3] &=& -e^{\im\xi}(D_1\phi_{\bar 4} + \im[\chi_1,\phi_{\bar 4}]) \\
\label{eq:5.3.4}
\ast D\ast\chi - \frac{\chi_1}{R_1} - \frac{\chi_2}{R_2} &=& 2\im[\phi_i,\phi_{\bar i}] \ ,
\end{eqnarray}
on the torus $T^2$.  Again, the constraint (\ref{Phi}) makes no restriction on the fields $\phi_i$.

\begin{figure}[ht]
\centerline{
\includegraphics[width=5cm]{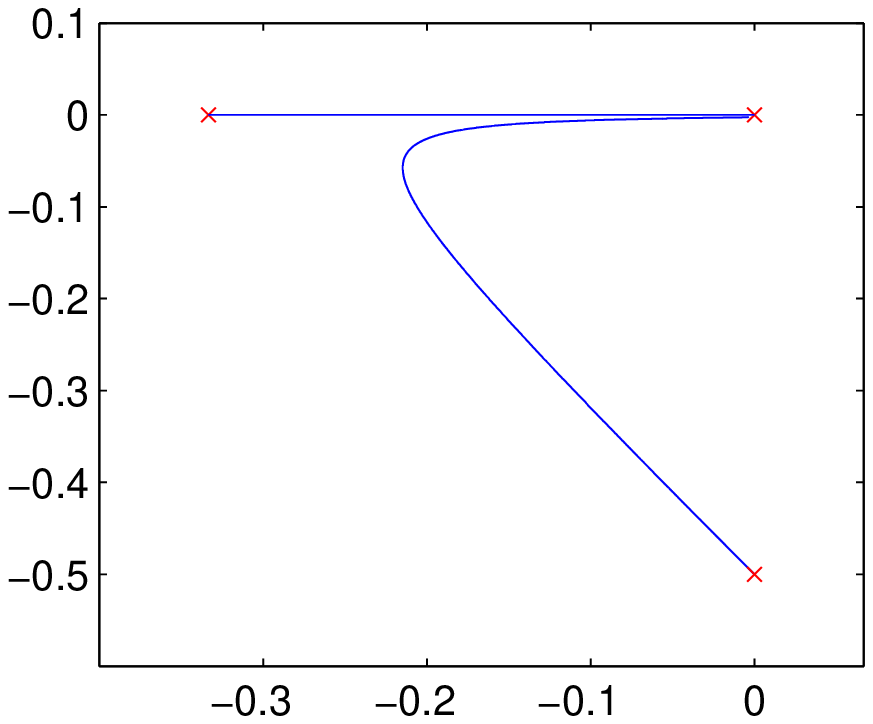}
%\hfill
\includegraphics[width=5cm]{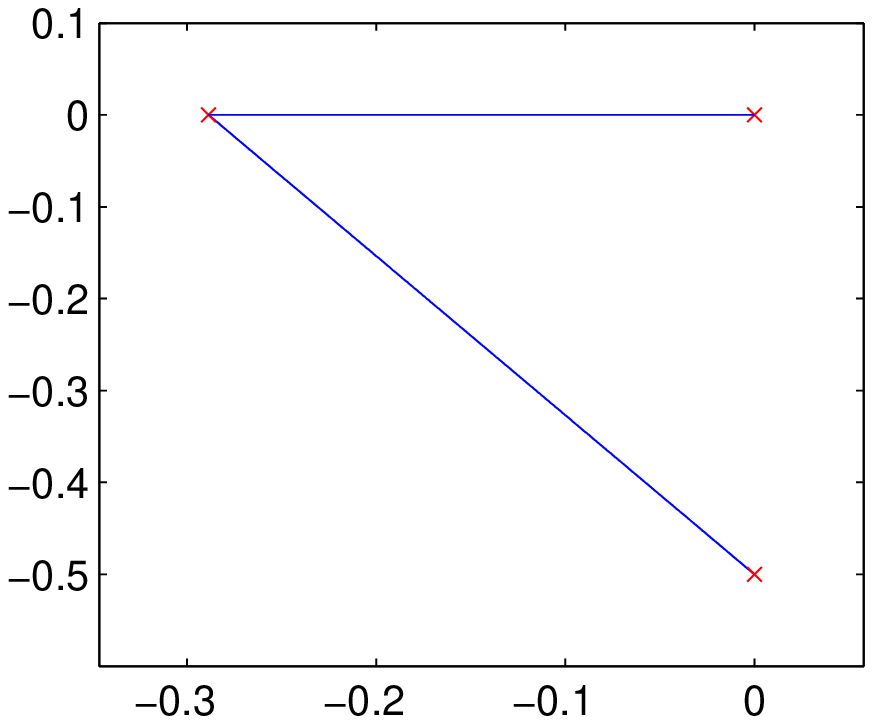}
%\hfill
\includegraphics[width=5cm]{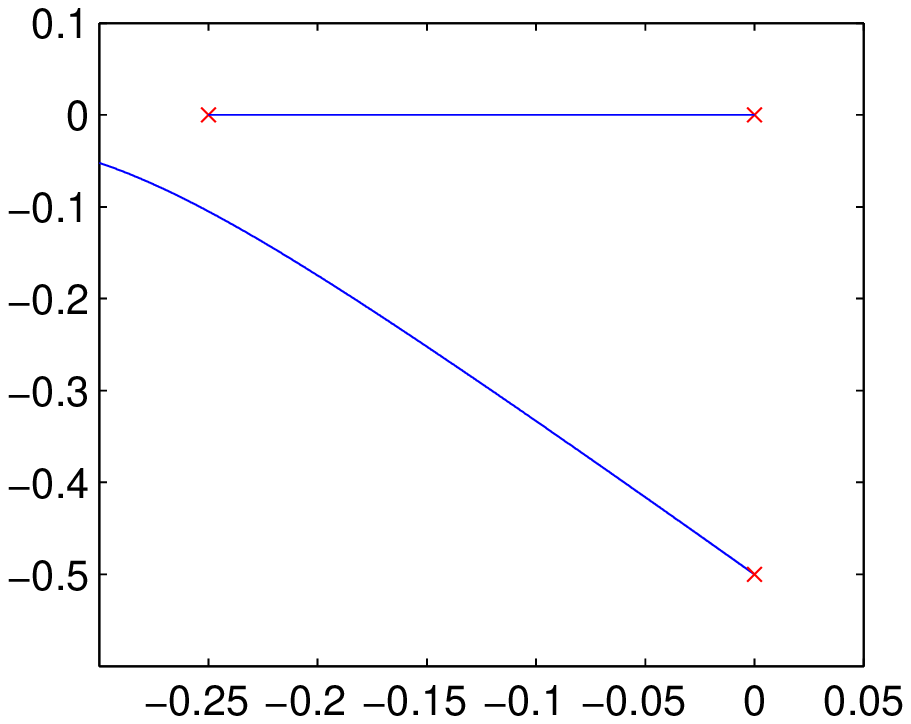}
}
\caption{Solutions of the flow equation in the plane spanned by $\psi$ (horizontal axis) and $\eta$ (vertical axis) with $R_2=1$ and $R_1=1.5$ (left), $R_1=\sqrt{3}\approx1.73$ (centre), and $R_1=2$ (right)}
\label{fig:1}
\end{figure}

Equations (\ref{eq:5.3.1})-(\ref{eq:5.3.4}) are nonlinear partial differential equations, and seems hard to find genuine 2-dimensional solutions.  However, we have found interesting 1-dimensional solutions.  In order to construct solutions, we make the ansatz,
\begin{equation}
\label{eq:5.3.5}
\phi_3 = \psi\sfrac{\im}{2}(\sigma_1-\im\sigma_2)\ ,\quad \phi_4 = \eta\sfrac{\im}{2}(\sigma_1+\im\sigma_2)\ ,\quad \chi_1 = \psi\im\sigma_3\ ,\quad \chi_2 = -\eta\im\sigma_3\ ,
\end{equation}
and set $\xi=0$, $\pa_2=0$ and $A_1=A_2=0$.  The Spin(7)-instanton equations simplify to
\begin{eqnarray}
\label{eq:5.3.6}
\partial_1\eta &=& -4\psi\eta \\
\label{eq:5.3.7}
\partial_1\psi &=& \frac{1}{R_1}\psi-\frac{1}{R_2}\eta + 2(\psi^2-\eta^2)\ .
\end{eqnarray}
These flow equations have three fixed points $P_i$: 
\begin{equation}
P_1: (\psi,\eta)=(0,0)\ ,\quad P_2: (\psi,\eta)=(-1/(2R_1),0)\ ,\quad P_3: (\psi,\eta)=(0,-1/(2R_2))\ .
\end{equation}
The qualitative features of the solutions of (\ref{eq:5.3.6}), (\ref{eq:5.3.7}) depend on the value of the ratio $\rho=R_1/R_2$.  For all $\rho$, there is a solution parametrised by a real constant $x^1_0$,
\begin{equation}
(\psi,\eta) = \left( 1-\exp\left(-\frac{1}{R_1}(x^1-x^1_0)\right)\right)^{-1} \left(-\frac{1}{2R_1},0\right) \ ,
\end{equation}
which asymptotes to $P_1$ at $x^1=-\infty$ and $P_2$ at $x^1=\infty$.  This in fact the well-known solution of the anti-self-dual Yang-Mills equations on $S^3\times\R$, lifted to $S^3\times S^3\times S^1\times\R$.  When $\rho=\sqrt{3}$, there is also an analytic solution,
\begin{equation}
(\psi,\eta) = \left(-\frac{1}{2R_1},0\right) + \frac{1}{2R_1} \left( 1+\exp\left(-\frac{2}{R_1}(x^1-x^1_0)\right)\right)^{-1} \left(1,-\frac{1}{\sqrt{3}}\right)\ ,
\end{equation}
which asymptotes to $P_2$ at $x^1=-\infty$ and $P_3$ at $x^1=\infty$.  For other values of $\rho$ we have constructed solutions numerically, see figure \ref{fig:1}.  For $\rho<\sqrt{3}$ there exists a solution asymptoting to $P_1$ at $x^1=-\infty$ and $P_3$ at $x^1=\infty$, but for $\rho\geq\sqrt{3}$ this solution ceases to exist.  Analysis of the critical points $P_1,P_2,P_3$ tells us that no other solutions asymptoting to the $P_i$ at $x^1=\pm\infty$ exist.

As a point of interest, our ansatz (\ref{eq:5.3.5}) has implicitly imposed invariance of $\Acal$ under the right action of SU(2).  It is surprising that the equations are consistent with this ansatz, since the SU(4)-structure (\ref{eq:5.3.0}) is not invariant under the right action of SU(2).  The explanation is that there is another SU(4)-structure which is invariant under the right action of SU(2):
\begin{equation}
\tilde\Th^1 = \hat e^1+\im\check e^1 ,\quad \tilde\Th^2 = \hat e^2-\im\check e^2 ,\quad \tilde\Th^3 = \hat e^3-\im\check e^3 ,\quad \tilde\Th^4 = \diff x^1+\im\diff x^2 \ .
\end{equation}

We denote by $\omega$ and $\tilde\omega$ the (1,1)-forms determined by the frames $\Theta^\mu$, $\tilde\Theta^\mu$.  Then
\begin{eqnarray}
\tilde\omega &=& {\rm Re}(\Th^1\wedge\Th^2+\Th^{\bar 3}\wedge\Th^{\bar 4})\ ,\\
\omega &=& {\rm Re}(\tilde\Theta^1\wedge\tilde\Theta^2 - \tilde\Theta^{\bar 3}\wedge\tilde\Theta^{\bar4})\ .
\end{eqnarray}
So $\tilde\omega$ solves the complex self-dual equation (\ref{complex SD}) for the frame $\Theta^\mu$ with $\xi=0$, and $\omega$ solves the complex self-dual equation for the frame $\tilde\Theta^\mu$ with $\xi=\pi$.  Therefore the Spin(7)-instanton equations for the frame $\Theta^\mu$ with $\xi=0$ are equivalent to the Spin(7)-instanton equations for the frame $\tilde\Theta^\mu$ with $\xi=\pi$.

\subsection{Invariant gauge fields on $S^3\times S^1\times T^4$}
\label{sec:5.4}

Now we turn our attention to $S^3\times T^5$.  Let $e^1,e^2,e^3$ be three left-invariant 1-forms on SU(2) as above, scaled so that $\diff e^a=-(1/2R)\eps_{abc}e^b\wedge e^c$, and let $e^4$ be an invariant
1-form on U(1).  We define an SU(4)-structure on $S^3\times S^1\times T^4$ by
\begin{equation}
\Th^1 = e^1+\im e^2\ ,\quad \Th^2=e^3+\im e^4\ , \quad \Th^3=\diff z^1\ , \quad \Th^4 = \diff z^2\ ,
\end{equation}
with $z^1,z^2$ complex coordinates on $T^4$.  This structure is CYT, because the canonical torsion is
\begin{equation}
T=-\frac{1}{R}\,e^1\wedge e^2\wedge e^3 =
-\frac{\im}{4R}\,\Th^1\wedge\Th^{\bar1}\wedge(\Th^2+\Th^{\bar2})\ .
\end{equation}

A gauge field on $T^4\times S^3\times S^1$ invariant under the left action
of ${\rm SU}(2)\times{\rm U}(1)$ may be written using the ansatz (\ref{coset gauge field}), with $A$ a gauge field on $T^4$ taking values in $\mathfrak{k}$ and
\begin{equation}
\Phi = \phi_i\Th^i + \phi_{\bar i}\Th^{\bar i}\ ,\quad i=1,2.\
\end{equation}
Here $\phi_i$ are two scalars on $T^4$ taking values in $\mathfrak{k}_\C$.
Then the Spin(7)-instanton equations for $\Fcal$ on $S^3\times S^1\times T^4$ reduce to the equations,
\begin{eqnarray}
\label{S3xS1 1}
e^{\im\xi}D_{\bar2}\phi_{\bar2} &=& D_1 \phi_1\ , \\
\label{S3xS1 2}
-e^{\im\xi}D_{\bar2}\phi_{\bar1} &=& D_1\phi_2\ , \\
\label{S3xS1 3}
-e^{\im\xi}F_{\bar1\bar2} &=& \frac{\im}{2R} \phi_1 +[\phi_1,\phi_2]\ , \\
\label{S3xS1 4}
F_{1\bar1} + F_{2\bar2} &=& -[\phi_1,\phi_{\bar1}]-[\phi_2,\phi_{\bar2}]+
\frac{\im}{2R}(\phi_2+\phi_{\bar2})\ ,
\end{eqnarray}
on $T^4$ (or one of its decompactifications $T^{4-p}\times \R^p$).

\subsection{Kapustin-Witten equations}

An alternative way to carry out the reduction from section \ref{sec:5.4} is to pick a different
SU(4)-structure compatible with the same Spin(7)-structure.  Consider again $S^3\times S^1\times T^4$ (or more generally, $S^3\times S^1\times T^{4-p} \times \R^p$).
Let $x^\mu$ be real coordinates on $T^4$, so that $z^1=x^1-\im x^2$ and $z^2=x^3-\im x^4$, and let
\begin{equation}
\tilde\Th^\mu = \diff x^\mu + \im e^\mu\ .
\end{equation}
We denote by $\omega$ and $\tilde\omega$ the (1,1)-forms determined by the frames $\Theta^\mu$, $\tilde\Theta^\mu$.  Then
\begin{eqnarray}
\tilde\omega &=& -{\rm Re}(\Theta^1\wedge\Theta^3 + \Theta^{\bar2}\wedge\Theta^{\bar4})\ ,\\
\omega &=& -{\rm Re}(\tilde\Theta^1\wedge\tilde\Theta^2 + \tilde\Theta^{\bar3}\wedge\tilde\Theta^{\bar4})\ .
\end{eqnarray}
So $\tilde\omega$ solves the complex self-dual equation (\ref{complex SD}) for the frame $\Theta^\mu$ with $\xi=\pi$, and $\omega$ solves the complex self-dual equation for the frame $\tilde\Theta^\mu$ with $\xi=0$.  Therefore the Spin(7)-instanton equations for the frame $\tilde\Theta^\mu$ with $\xi=0$ are equivalent to the Spin(7)-instanton equations for the frame $\Theta^\mu$ with $\xi=\pi$.
With respect to the frame $\tilde\Theta^\mu$, the canonical torsion is still a 3-form, but is neither nearly
K\"ahler nor CYT:
\begin{equation}
T=-\frac{1}{R}e^1\wedge e^2\wedge e^3 = \frac{\im}{8R}(\tilde\Th^{\bar1}-
\tilde\Th^1)\wedge(\tilde\Th^{\bar2}-\tilde\Th^2)\wedge(\tilde\Th^{\bar3}-
\tilde\Th^3)\ .
\end{equation}

We make the same ansatz (\ref{coset gauge field}) for the gauge field, but parametrise $\Phi$ in terms of four $\mathfrak{k}$-valued scalars $\chi_\mu$ rather than the two $\mathfrak{k}_\C$-valued scalars $\phi_1,\phi_2$:
\begin{equation}
\Phi = \chi_\mu e^\mu\ ,\quad \mu=1,\dots,4\ .
\end{equation}
If we write $\chi=\chi_\mu\diff x^\mu$, the complex anti-self-dual
equation (\ref{SU4inst1}) is reduced to the following equations in 4 dimensions:
\begin{eqnarray}
\label{S3xS1 5}
\cos\frac{\xi}{2}\left(F-\chi\wedge\chi +
\frac{1}{R} \ast(\chi\wedge \diff x^4)\right) -
\sin\frac{\xi}{2} D \chi && \mbox{ is anti-self-dual}\ , \\
\label{S3xS1 6}
\cos\frac{\xi}{2} D \chi + \sin\frac{\xi}{2}\left(F-\chi\wedge\chi +
\frac{1}{R} \ast(\chi\wedge \diff x^4)\right) && \mbox{ is self-dual}\ .
\end{eqnarray}
The second of the Spin(7)-instanton equations (\ref{SU4inst2}) becomes
\begin{equation}
\label{S3xS1 7}
D \ast \chi = 0\ .
\end{equation}
Equations (\ref{S3xS1 5})-(\ref{S3xS1 7}) with $\xi=0$ are equivalent to equations (\ref{S3xS1 1})-(\ref{S3xS1 4}) with $\xi=\pi$.

The Hermitian-Yang-Mills equations are slightly stronger: they consist of (\ref{S3xS1 7}), and the vanishing of the (2,0) part of $\Fcal$:
\begin{eqnarray}
\label{S3xS1 8}
\left(F-\chi\wedge\chi +
\frac{1}{R} \ast(\chi\wedge \diff x^4)\right) &=&0\ , \\
\label{S3xS1 9}
D \chi&=&0\ .
\end{eqnarray}
Note that equations (\ref{S3xS1 5})-(\ref{S3xS1 9}) on $\R^4$ or $T^4$ agree with the Kapustin-Witten equations (\ref{KW1})-(\ref{KW5}) in the limit $R\rightarrow\infty$.  So these equations are a simple,
geometrically-motivated perturbation of Kapustin and Witten's -- it would
be interesting to see whether a topological field theory can be
developed from them.

The Kapustin-Witten equations (\ref{KW1})-(\ref{KW3}) only have non-trivial solutions in the cases $t=0,\infty$.  We have not found any solutions to our equations (\ref{S3xS1 5})-(\ref{S3xS1 7}) on $T^4$ or $T^{4-p}\times \R^p$ (besides the usual Yang-Mills instantons); however, the Spin(7)-instanton constructed on $S^3\times S^3\times S^1\times\R$ in section \ref{sec:5.3} corresponds to a solution of (\ref{S3xS1 5})-(\ref{S3xS1 7}) on $S^3\times\R$ invariant under rotations of $S^3$.

\section{Yang-Mills fields on the ${\rm SU}(2)$-structure CYT spaces}
\label{sec:6}

In this section we discuss the Spin(7)-instanton equations on two CYT coset spaces whose natural structure group is SU(2).

\subsection{Invariant gauge fields on $(S^5\times S^1)\times T^2$}

The coset space ${\rm SU}(3)/{\rm SU}(2)$ is $S^5$ as a manifold.  A local basis of 1-forms $e^a$ can be constructed from generators $I_a$ as in section \ref{sec:4}.  We choose the quadratic form 
\begin{equation}
q(X,Y) = -\frac{R^2}{4}\langle X,Y\rangle_{CK} = -\frac{3R^2}{2}{\rm Tr}_3(XY)
\end{equation}
on $\mathfrak{su}(3)$, with $\langle\cdot,\cdot\rangle_{CK}$ denoting the Cartan-Killing form.  An orthonormal basis for $\mathfrak{m}$ is
\begin{equation}
\label{basis su3}
\begin{aligned}
I_1=&\frac{1}{\sqrt3R}
\left(\begin{array}{ccc}0&-1&0\\1&0&0\\0&0&0\end{array}\right), &
I_3=&\frac{1}{\sqrt3R}
\left(\begin{array}{ccc}0&0&-1\\0&0&0\\1&0&0\end{array}\right), &
I_5=&\frac{1}{3R}
\left(\begin{array}{ccc}2\im&0&0\\0&-\im&0\\0&0&-\im\end{array}\right), \\
I_2=&\frac{1}{\sqrt3R}
\left(\begin{array}{ccc}0&\im&0\\ \im&0&0\\0&0&0\end{array}\right) , &
I_4=&\frac{1}{\sqrt3R}
\left(\begin{array}{ccc}0&0&\im\\0&0&0\\ \im&0&0\end{array}\right)\ . & &
\end{aligned}
\end{equation}
The corresponding local frame $e^a$ on $S^5$ is orthornormal with respect to the metric induced by $q$.

As explained in section \ref{sec:4}, there is a natural SU(2)-structure on $S^5={\rm SU}(3)/{\rm SU}(2)$.  This SU(2)-structure can be lifted to an SU(4)-structure on $S^5\times S^1\times T^2$, defined by
\begin{equation}
\Th^1=e^1+\im e^2,\quad \Th^2=e^3+\im e^4,\quad \Th^3=e^5+\im\diff t,\quad \Th^4 = \diff z,
\end{equation}
with $t$ a coordinate on $S^1$ and $z$ a complex coordinate on $T^2$.  The torsion (\ref{torsion}) of this SU(4)-structure is
\begin{equation}
T = \frac{1}{R}(e^1\wedge e^2+e^3\wedge e^4)\wedge e^5 =
\frac{\im}{4R}(\Th^1\wedge\Th^{\1}+\Th^2\wedge\Th^{\2})\wedge(\Th^3+
\Th^{\3})\ .
\end{equation}
Since $T$ is the real part of a (2,1)-form, the manifold is CYT.
It is worth stressing that the metric on $S^5$ which makes $S^5\times S^1$
CYT is not the usual round metric.

We make the standard ansatz (\ref{coset gauge field}) for an SU(3)-invariant gauge field, with
\begin{equation}
\Phi = \phi_i\Th^i + \phi_{\bar i}\Th^{\bar i} + \chi e^5, \quad i=1,2 \ ,
\end{equation}
where $\phi_i$ take values in $\mathfrak{k}_\C$ and $\chi$ take values in $\mathfrak{k}$, and both are subject to the constraint (\ref{Phi}).  The Spin(7)-instanton equations on $S^5\times S^1\times T^2$ for the field strength
(\ref{coset field strength}) reduce to monopole-type equations,
\begin{eqnarray}
\im F_{zt}-D_z\chi &=& -2e^{\im\xi}[\phi_{\bar 1},\phi_{\bar 2}]\ , \\
2D_z\phi_2 &=& e^{\im\xi}\left(\im D_t\phi_{\bar 1}+[\chi,\phi_{\bar 1}]-\frac{\im}{2R}\phi_{\bar 1}\right)\ , \\
2D_z\phi_1 &=& -e^{\im\xi}\left(\im D_t\phi_{\bar 2}+[\chi,\phi_{\bar 2}]-\frac{\im}{2R}\phi_{\bar 2}\right)\ , \\
2\im F_{z\bar z}+D_t\chi-\frac{2}{R}\chi &=& -2\im[\phi_i,\phi_{\bar i}]\ ,
\end{eqnarray}
on $T^3$ (or one of its decompactifications $\T^{3-p}\times\R^p$).

In principle one could choose any gauge group $K$, but to give a concrete example we make the choice $K={\rm SU}(3)$, and take the homomorphism from SU(2) to SU(3) to be the standard inclusion.  The reduced gauge group is U(1), the centraliser of SU(2) in SU(3), and we set $A=RaI_5$,
with $a$ a real 1-form.  The constraint (\ref{Phi}) is solved by
\begin{equation}
\phi_i=\phi RY_i\ ,\quad \chi = \psi RI_5 \, ,
\end{equation}
with $\phi,\psi$ two complex scalars and $Y_i:=\sfrac12 (I_{2i-1}-\im I_{2i})$.  With these choices, the
Spin(7)-instanton equations reduce to
\begin{eqnarray}
(\diff a)_{zt} + \im\partial_z\psi &=& 0\\
\pa_z\phi - \im a_z\phi &=&0\\
\pa_t \phi-\im a_t \phi+ \left( \psi-\frac{1}{R}\right)\phi &=&0\\
2\im(\diff a)_{z\bar z} + 2|\phi|^2+\pa_t\psi-\frac{2}{R}\psi &=&0.
\end{eqnarray}
These equations imply not just the Spin(7)-instanton equations, but also the Hermitian-Yang-Mills equations.  To find solutions, we assume $t$-invariance and write $\eta=\psi-\im a_t$.  The equations reduce further to
\begin{eqnarray}
\partial_{z}\bar\phi - \im a_{z}\bar\phi &=& 0\ , \\
\partial_{z}\bar\eta &=& 0\ , \\
\phi\left(\eta-\frac{1}{R}\right) &=& 0\ , \\
\ast \diff a &=& 2|\phi|^2-\frac{2}{R}{\rm Re}(\eta)\ .
\end{eqnarray}
These equations force either $\phi=0$ or $\eta=1/R$; the latter case is equivalent to the Bogmolny equation of the abelian Higgs model, which is known to admit many interesting solutions on both $\R^2$ and $T^2$.

\subsection{Invariant gauge fields on $S^5\times S^3$}

As above, we introduce a local frame of 1-forms on $S^5$, now denoted $\hat e^a$ with $a=1,\dots,5$.  We denote by $\check e^a$, $a=1,2,3$, the left-invariant 1-forms on SU(2) satisfying $\diff \check e^a=(1/2\Lambda)\epsilon_{abc}\check e^b\wedge \check e^c$.  An SU(4)-structure on $S^5\times S^3$ is given by,
\begin{equation}
\Th^1=\hat e^1+\im\hat e^2,\quad \Th^2=\hat e^3+\im\hat e^4,\quad \Th^3=\check e^1+\im\check e^2,\quad \Th^4=\hat e^5+\im\check e^3\ ,
\end{equation}
and this lifts the natural SU(2)-structure described in section \ref{sec:4}.  The canonical torsion is the real part of a (2,1)-form, so the SU(4)-structure is CYT:
\begin{eqnarray}
T &=& \frac{1}{R}(\hat e^1\wedge\hat e^2 +\hat e^3\wedge\hat e^4)\wedge\hat e^5 + \frac{1}{\La}\check e^1\wedge\check e^2\wedge\check e^3 \\
&=& \frac{\im}{4R}(\Th^1\wedge\Th^{\bar 1} + \Th^2\wedge\Th^{\bar 2})\wedge(\Th^4+\Th^{\bar 4}) + \frac{1}{4\La}\Th^3\wedge\Th^{\bar 3}\wedge(\Th^4-\Th^{\bar 4}).
\end{eqnarray}

We make the usual ansatz (\ref{coset gauge field}) for the gauge field, with $A=0$ and
\begin{equation}
\Phi = \phi_\mu \Th^\mu + \phi_{\bar\mu}\Th^{\bar\mu}\ ,\quad \mu=1,\dots,4\ ,
\end{equation}
where $\phi_\mu$ take values in $\mathfrak{k}_\C$ and are subject to the constraint (\ref{Phi}).  The Spin(7)-instanton equations are reduced to matrix model equations in $d=0$ dimensions:
\begin{eqnarray}
e^{\im\xi}[\phi_{\bar1},\phi_{\bar2}] &=& [\phi_4,\phi_3] + \sfrac{1}{2\La}\phi_3\ , \\
e^{\im\xi}[\phi_{\bar2},\phi_{\bar3}] &=& [\phi_4,\phi_1] + \sfrac{\im}{2R}\phi_1\ , \\
e^{\im\xi}[\phi_{\bar3},\phi_{\bar1}] &=& [\phi_4,\phi_2] + \sfrac{\im}{2R}\phi_3\ , \\
2\im\sum_{\mu=1}^4[\phi_\mu,\phi_{\bar\mu}] &=& (\sfrac{2}{R}-\sfrac{\im}{\La})\phi_4 + (\sfrac{2}{R}+\sfrac{\im}{\La})\phi_{\bar 4}\ .
\end{eqnarray}

We choose gauge group $K={\rm SU}(3)$, with ${\rm SU}(2)\subset{\rm SU}(3)$ embedded in the obvious way.  The constraint (\ref{Phi}) is solved by
\begin{equation}
\phi_1 = \phi Y_1, \quad \phi_2=\phi Y_2,\quad \phi_3 = \psi I_7,\quad \phi_4 = \chi I_7,
\end{equation}
for complex $\phi,\chi,\psi$.  The Spin(7)-instanton equations are become $\psi=0$ and
\begin{eqnarray}
\phi(\chi-\sfrac{1}{2}) &=& 0 \\
|\phi|^2 &=& {\rm Re}((2-\im\sfrac{R}{\La})\chi).
\end{eqnarray}
Once again, these equations imply the Hermitian-Yang-Mills equations as well as the Spin(7)-instanton equations.  They have two types of solution: either $\phi=0$ and $\chi=\lambda(R/\Lambda-2\im)$ for some $\lambda\in\R$; or $\chi=\sfrac12$ and $|\phi|^2=1$ (the canonical connection $\phi=0,\chi=0$ is included in the former).

\section{Yang-Mills fields on SU(3)-structure CYT spaces}
\label{sec:7}

In this section, we study the Spin(7)-instanton equations on the CYT manifold $S^7\times S^1$.  As a manifold, $S^7$ is the coset space SU(4)/SU(3), so the natural structure group is SU(3).  Following the formalism described in section \ref{sec:4}, we choose a quadratic form on $\mathfrak{su}(4)$:
\begin{equation}
q(X,Y) = -\frac{R^2}{6}\langle X,Y\rangle_{CK} = -\frac{4R^2}{3}{\rm Tr}_4(XY).
\end{equation}
An orthonormal basis for $\mathfrak{m}$ is
\begin{equation}
\begin{aligned}
I_1=&\frac{\sqrt{6}}{4R}\left(\begin{array}{cccc}0&-1&0&0\\1&0&0&0\\0&0&0&0\\0&0&0&0\end{array}\right),&
I_3=&\frac{\sqrt{6}}{4R}\left(\begin{array}{cccc}0&0&-1&0\\0&0&0&0\\1&0&0&0\\0&0&0&0\end{array}\right),&
I_5=&\frac{\sqrt{6}}{4R}\left(\begin{array}{cccc}0&0&0&-1\\0&0&0&0\\0&0&0&0\\1&0&0&0\end{array}\right),\\
I_2=&\frac{\sqrt{6}}{4R}\left(\begin{array}{cccc}0&\im&0&0\\\im&0&0&0\\0&0&0&0\\0&0&0&0\end{array}\right),&
I_4=&\frac{\sqrt{6}}{4R}\left(\begin{array}{cccc}0&0&\im&0\\0&0&0&0\\\im&0&0&0\\0&0&0&0\end{array}\right),&
I_6=&\frac{\sqrt{6}}{4R}\left(\begin{array}{cccc}0&0&0&\im\\0&0&0&0\\0&0&0&0\\\im&0&0&0\end{array}\right),\\
I_7=&\frac{1}{4R}\left(\begin{array}{cccc}3\im&0&0&0\\0&-\im&0&0\\0&0&-\im&0\\0&0&0&-\im\end{array}\right).
\end{aligned}
\end{equation}
This induces a local frame $e^1,\dots,e^7$ of 1-forms.  An SU(4)-structure on $S^7\times S^1$ which lifts the natural SU(3)-structure is induced by
\begin{equation}
\Th^\a=e^{2\a-1}+\im e^{2\a}\ ,\quad \Th^4 = e^7+\im \diff t\ ,
\end{equation}
with $\a=1,2,3$ and $t$ a coordinate on $S^1$.  The canonical torsion is the real part of a (2,1)-form, so the SU(4)-structure is CYT:
\begin{equation}
T = \frac{1}{R}(e^1\wedge e^2+e^3\wedge e^4+ e^5\wedge e^6)\wedge e^7 = \frac{\im}{4R}\Th^\a\wedge\Th^{\bar\a}\wedge(\Th^4+\Th^{\bar4})\ .
\end{equation}

We make the usual ansatz for an SU(4)-invariant gauge field (\ref{coset gauge field}), with
\begin{equation}
\Phi=\phi_\a\Th^\a+\phi_{\bar\a}\Th^{\bar\a}+\chi e^7\ ,\quad\a=1,2,3\ ,
\end{equation}
and the gauge chosen so that $A=0$.  The scalar $\chi$ takes values in the Lie algebra $\mathfrak{k}$ of $K$ and the scalars $\phi_\a$ take values in $\mathfrak{k}_\C$; together they must satisfy the constraint (\ref{Phi}).  The Spin(7)-instanton equations are then reduced to the equations,
\begin{eqnarray}
\frac{\im}{2}\left(\dot\phi_\a-\frac{1}{R}\phi_\a+\im[\chi,\phi_\a]\right) &=& -\frac{1}{2}e^{\im\xi}\eps_{\a\b\g} [\phi_{\bar\b},\phi_{\bar\g}]\ ,\\
\dot\chi-\frac{3}{R}\chi&=&-2\im[\phi_\a,\phi_{\bar\a}]\ ,
\end{eqnarray}
on $S^1$ (or its decompactification $\R$).  We choose gauge group SU(4) with SU(3) embedded in the obvious way, so that the constraint (\ref{Phi}) is solved by
\begin{equation}
\phi_\a=\phi RY_\a\ ,\quad \chi=\psi RI_7\ ,\quad Y_\a=\sfrac12(I_{2\a-1}-\im I_{2\a}).
\end{equation}
Then the Spin(7)-instanton equations reduce to
\begin{eqnarray}
\dot\phi &=& \phi\left( \frac{1}{R} - \psi\right)\\
\dot\psi &=& 3\left( \frac{1}{R}\psi - | \phi |^2 \right).
\end{eqnarray}

These equations imply the Hermitian-Yang-Mills equations as well as the Spin(7)-instanton equations.  They are gradient flow equations for a function $W$:
\begin{eqnarray}
\dot\phi &=& \frac{\pa W}{\pa\bar\phi}\\
\dot\psi &=& 3\frac{\pa W}{\pa\psi}\\
W &=& \frac{1}{R}| \phi |^2+\frac{1}{2R}\psi^2-\psi | \phi |^2.
\end{eqnarray}
Since the value of $W$ decreases along the flow, the only solutions with $t\in S^1$ are $t$-independent: either $\psi=0,\phi=0$ (the canonical connection) or $\psi=1/R,|\phi|=1/R$.  It is also interesting to look for solutions with $t\in\R$ which interpolate between the critical points of $W$.  The critical point $\psi=0,\phi=0$ is a local minimum for $W$, and the critical point $\psi=1/R,|\phi|=1/R$ is a saddle point; it follows that there is a unique (up to translation) solution which tends to $\psi=1/R,|\phi|=1/R$ as $t\to-\infty$ and to $\psi=0,\phi=0$ as $t\to\infty$.

\section{Yang-Mills fields on nearly K\"ahler manifolds and vortex equations}
\label{sec:8}

\subsection{Reduced Spin(7)-instanton equations}
\label{sec:gen nk}

There are only four known examples of nearly K\"ahler 6-manifolds, and they
are all cosets \cite{But}: 
\begin{equation}
\begin{array}{cc}
{\rm SU}(3)/{\rm U}(1){\times}{\rm U}(1)\ , & {\rm Sp}(2)/{\rm Sp}(1){\times}{\rm U}(1)\ ,\\
G_2/{\rm SU}(3)=S^6\ , & {\rm SU}(2)^3/{\rm SU}(2)=S^3\times S^3\ .
\end{array}
\end{equation}
We will consider the dimensional reduction of the Spin(7)-instanton equations over these spaces.

The SU(3)-structure on the coset space is determined in an elegant way from the
Lie algebra $\mathfrak{g}$.  For example, the metric $g$ is determined by
the Cartan-Killing form, and the complex structure $J$ is related to the
3-symmetry, a linear map $S:\mathfrak{m}\rightarrow\mathfrak{m}$ satisfying
$S^3=1$.  Details of this construction can be found in \cite{HILP,But}.

Here we will specify the nearly K\"ahler SU(3)-structures on the coset spaces more directly.  We choose the quadratic form,
\begin{equation}
q(X,Y) = -\frac{R^2}{12}\langle X,Y\rangle_{CK}\ ,
\end{equation}
on all four coset spaces.  For each coset space, we will write down an orthonormal basis $I_a$ for $\mathfrak{m}$.  Then left-invariant 1-forms $e^a$ are induced on the coset spaces as described in section \ref{sec:4}, and we can define
\begin{equation}
\label{NK structure}
\Th^\a = e^{2\a-1} + \im e^{2\a} \ , \quad \a = 1,2,3\ .
\end{equation}
The SU(3)-structure is then determined by $J,\ome,\Om$ written in the standard form (\ref{SU3 structure}).  An SU(4) structure on $G/H\times T^2$ is defined by setting $\Th^4=\diff z$, with $z$ a complex coordinate on $T^2$.  It will prove convenient to define $Y_\a=\sfrac12(I_{2\a-1}-\im I_{2\a})$.  In section \ref{sec:SU(2)^3/SU(2) alt} we will use different conventions.

The four nearly K\"ahler coset spaces are algebraically very similar; for example, the torsion tensor (\ref{torsion}) is
\begin{equation}
T = -\frac{1}{6}\, f_{abc}\, e^a\wedge e^b\wedge e^c =
\frac{1}{R}\ {\rm Re}\,\Om
\end{equation}
This means that the reduction of
Spin(7)-instanton equations depends on the choice of nearly K\"ahler coset space
only through the constraint (\ref{Phi}).

We will make the usual ansatz for a $G$-invariant gauge field (\ref{coset gauge field}), with $\Phi=\phi_\a\Th^\a +
\phi_{\bar\a}\Th^{\bar\a}$ and $\phi_\a$ scalars on $T^2$ taking values in the complexification of the Lie algebra of $K$.
The Spin(7)-instanton equations for $\Fcal$ are then reduced to the equations,
\begin{eqnarray}
\label{NK inst 1}
e^{\im\xi}D_{\bar{z}}\phi_{\bar{\a}} &=&
\sfrac{1}{2}\,\epsilon_{\bar\a\bar\b\bar\g}[\phi_\b,\phi_\g] +
\sfrac{1}{R}\, \phi_{\bar \a}\ , \\
\label{NK inst 2}
\ast F &=& 2\im [\phi_\a,\phi_{\bar{\a}}] \ ,
\end{eqnarray}
on $T^2$.  The field strength $F^0$ of the canonical connection solves the
Spin(7)-instanton equations on a nearly K\"ahler coset space, so it does not appear in these
equations.  Thus the only differences between these equations and
(\ref{gen hitchin 1}), (\ref{gen hitchin 2}) come from the torsion term in
(\ref{coset field strength}) and the constraint (\ref{Phi}) on the
fields $\phi_\a$.

The real metric $q$ on $\mathfrak{g}$ can be extended to define a Hermitian metric on $\mathfrak{g}_\C$.  Then, similar to equation (\ref{gen hitchin 1}), equation (\ref{NK inst 1}) is the Hermitian flow equation for the function
\begin{equation}
W(\phi_1,\phi_2,\phi_3) = q( \phi_1,[\phi_2,\phi_3]) +q(\phi_{\bar1},[\phi_{\bar2},\phi_{\bar3}]) + \frac{1}{R}q(\phi_\a,\phi_{\bar\a})\ .
\end{equation}

Now we will give examples of the Spin(7)-instanton equations on the SU(4)-structure manifold $T^2\times G/H$ for each of
the four nearly K\"ahler cosets $G/H$.

\subsection{Invariant gauge fields on $T^2\times G_2/{\rm SU}(3)$}

This case was discussed in \cite{HILP}.  An obvious choice of gauge group is $G_2$, and a bundle over $S^6$ admitting an action of $G_2$ is determined by the obvious embedding of SU(3) in $G_2$.  Then the reduced gauge group is trivial, and the solution of the constraint (\ref{Phi}) is parametrised by a complex scalar $\phi$.  The Spin(7)-instanton equations (\ref{NK inst 1}), (\ref{NK inst 2}) reduce to
\begin{equation}
\label{eq:8.1.1}
e^{\im\xi}\partial_{\bar z}\bar\phi = \frac{1}{R}(\bar\phi-\phi^2)\ .
\end{equation}

\subsection{Invariant gauge fields on $T^2\times{\rm Sp}(2)/{\rm Sp}(1)\times{\rm U}(1)$}

We choose the following normalised basis for $\mathfrak{m}$
\begin{equation}
\begin{aligned}
I_1 =& \frac{1}{\sqrt{2}R}\left(\begin{array}{cc}0&-1_2\\1_2&0\end{array}\right), &
I_3 =& \frac{1}{\sqrt{2}R}\left(\begin{array}{cc}0&-\im\sigma^1\\-\im\sigma^1&0\end{array}\right),&
I_5 =& \frac{1}{R}\left(\begin{array}{cc}0&0\\0&\im\sigma^1\end{array}\right), \\
I_2 =& \frac{1}{\sqrt{2}R}\left(\begin{array}{cc}0&-\im\sigma^3\\-\im\sigma^3&0\end{array}\right), &
I_4 =& \frac{1}{\sqrt{2}R}\left(\begin{array}{cc}0&-\im\sigma^2\\-\im\sigma^2&0\end{array}\right), &
I_6 =& \frac{1}{R}\left(\begin{array}{cc}0&0\\0&-\im\sigma^2\end{array}\right)\ .
\end{aligned}
\end{equation}
The nearly K\"ahler SU(3)-structure on ${\rm Sp}(2)/{\rm Sp}(1)\times{\rm U}(1)$ is written in the standard form (\ref{NK structure}) in terms of the 1-forms $e^a$ induced by the $I_a$.

We choose gauge group $K={\rm Sp}(2)$ and take the homomorphism from
Sp(1)$\times$U(1) to be the inclusion.  The reduced gauge group is the
centraliser U(1) of Sp(1)$\times$U(1), and the gauge field on $T^2$ will
be written
\begin{equation}
A = a \left(\begin{array}{cc}0&0\\0&-\im\sigma^3\end{array}\right)
\end{equation}
with $a$ a real 1-form.  The solution of the condition (\ref{Phi}) is
\begin{equation}
\phi_1=\phi Y_1\ ,\quad
\phi_2=\phi Y_2\ ,\quad
\phi_3=\chi Y_3,
\end{equation}
with $\phi,\chi$ two complex scalars.

There is actually a 1-parameter
family of Spin(2)-invariant SU(3)-structures on the space ${\rm Sp}(2)/{\rm Sp}(1)\times{\rm U}(1)$, which includes
the nearly K\"ahler structure.  With respect to the fixed frame $\Th^\a$,
this family is written
\begin{equation}
\omega = \frac{{\im}}{2}\left( \Lambda^2\,\Th^1\wedge\bar\Th^1 +
\Lambda^2\,\Th^2\wedge\bar\Th^2 +
\Lambda^{-2}\Th^3\wedge\bar\Th^3 \right)\ , \quad
\Om = \Lambda\, \Th^1\wedge\Th^2\wedge\Th^3\ ,
\end{equation}
with $\Lambda$ a positive real parameter and $g$ fixed by $J,\ome$.  The 2-form $\omega$ is a genuine Sp(2)-invariant form on the coset, because it is induced from an ${\rm Sp}(1)\times{\rm U}(1)$-invariant form on $\mathfrak{m}$.

When $\Lambda=1$ the SU(3)-structure is nearly K\"ahler, and the reduction of the Spin(7)-instanton equations is obtained by substituting for $\phi_\a$ and $A$ in (\ref{NK inst 1}), (\ref{NK inst 2}).  For $\Lambda\neq1$ the reduction of the Spin(7)-instanton equations must be calculated directly from the formula (\ref{coset field strength}) for the field strength.  The result is:
\begin{eqnarray}
\label{eq:8.3.1}
(\pa_{\bar z}-\im a_{\bar z})\bar\phi &=&
\frac{\Lambda}{R}\,(\bar\phi-\phi\chi)\ , \\
\label{eq:8.3.2}
(\pa_{\bar z}+2\im a_{\bar z})\bar\chi &=&
\frac{1}{R\Lambda^3}\,(\bar\chi-\phi^2)\ , \\
\label{eq:8.3.3}
\ast \diff a &=& \frac{2}{R^2} \left( \frac{1}{\Lambda^2}-\Lambda^2-\frac{|\phi|^2}{\Lambda^2}+\Lambda^2|\chi|^2\right) \ .
\end{eqnarray}

\subsection{Invariant gauge fields on $T^2\times {\rm SU}(3)/{\rm U}(1)\times{\rm U}(1)$}

The normalised basis for $\mathfrak{m}$ which yields the nearly K\"ahler structure in standard form (\ref{NK structure}) is
\begin{equation}
\begin{aligned}
I_1=&\frac{1}{R}
\left(\begin{array}{ccc}0&-1&0\\1&0&0\\0&0&0\end{array}\right),&
I_3=&\frac{1}{R}
\left(\begin{array}{ccc}0&0&1\\0&0&0\\-1&0&0\end{array}\right),&
I_5=&\frac{1}{R}
\left(\begin{array}{ccc}0&0&0\\0&0&-1\\0&1&0\end{array}\right), \\
I_2=&\frac{1}{R}
\left(\begin{array}{ccc}0&\im&0\\ \im&0&0\\0&0&0\end{array}\right),&
I_4=&\frac{1}{R}
\left(\begin{array}{ccc}0&0&\im\\0&0&0\\ \im&0&0\end{array}\right),&
I_6=&\frac{1}{R}
\left(\begin{array}{ccc}0&0&0\\0&0&\im\\0&\im &0\end{array}\right).
\end{aligned}
\end{equation}

We choose gauge group $K={\rm SU}(3)$, with ${\rm U}(1)\times{\rm U}(1)$
embedded using the inclusion.  The condition (\ref{Phi}) is solved by
\begin{equation}
\phi_1 = \chi_1 Y_1\ ,\quad
\phi_2 = \chi_2 Y_2\ ,\quad
\phi_3 = \chi_3 Y_3\ ,
\end{equation}
with $\chi_\a$ complex scalars.  The reduced gauge group is
${\rm U}(1)\times{\rm U}(1)$, and we set
\begin{equation}
A = a\left(\begin{array}{ccc}\im&0&0\\0&-\im&0\\0&0&0\end{array}\right)
+ \frac{b}{3}\left(\begin{array}{ccc}-\im&0&0\\0&-\im&0\\0&0&2\im\end{array}\right)
\end{equation}
for real 1-forms $a,b$.

Once again, we may consider a family of SU(3)-structures parametrised by $\Lambda\in\R$, for which the nearly K\"ahler case appears when $\Lambda=1$:
\begin{equation}
\omega = \frac{{\im}}{2}\left( \Lambda^{-2}\,\Th^1\wedge\bar\Th^1 +
\Lambda^2\,\Th^2\wedge\bar\Th^2 +
\Lambda^2\Th^3\wedge\bar\Th^3 \right)\ , \quad
\Om = \Lambda\, \Th^1\wedge\Th^2\wedge\Th^3\ .
\end{equation}
The reduction of the Spin(7)-instanton equations is obtained by substituting for $\phi_\a$ and $A$ in the formula (\ref{coset field strength}) for the field strength, or in the case $\Lambda=1$ simply by substituting into (\ref{NK inst 1}), (\ref{NK inst 2}).  The result is:
\begin{eqnarray}
\label{eq:8.4.1}
e^{\im\xi}(\partial_{\bar z} +2\im a_{\bar z})\bar\chi_1 &=&
\frac{1}{R\Lambda^3}(\bar\chi_1-\chi_2\chi_3)\ , \\
\label{eq:8.4.2}
e^{\im\xi}(\partial_{\bar z} -\im a_{\bar z}+\im b_{\bar z})\bar\chi_2 &=&
\frac{\Lambda}{R}(\bar\chi_2-\chi_1\chi_3)\ , \\
\label{eq:8.4.3}
e^{\im\xi}(\partial_{\bar z} -\im a_{\bar z}-\im b_{\bar z})\bar\chi_3 &=&
\frac{\Lambda}{R}(\bar\chi_3-\chi_1\chi_2)\ , \\
\label{eq:8.4.4}
\ast\diff a &=& \frac{2}{R^2}\left(\frac{1}{\Lambda^2} - \Lambda^2- \frac{|\chi_2|^2+|\chi_3|^2}{2\Lambda^2} +\Lambda^2|\chi_1|^2\right)\ , \\
\label{eq:8.4.5}
\ast\diff b &=& \frac{3}{R^2\Lambda^2}(|\chi_2|^2-|\chi_3|^2)\ .
\end{eqnarray}

\subsection{Invariant gauge fields on $T^2\times {\rm SU}(2)^3/{\rm SU}(2)$}
\label{sec:SU(2)^3/SU(2) alt}

The Spin(7)-instanton equations can be reduced over ${\rm SU}(3)^3/{\rm SU}(2)$ using the formalism outlined in subsection \ref{sec:gen nk}, just as they can over the other three nearly K\"ahler coset spaces.  We will not do so here; instead, we reduce the equations using a different scheme which allows us to be more flexible about what symmetries are imposed.

We begin our discussion by observing that the coset space ${\rm SU}(3)^3/{\rm SU}(2)$ is diffeomorphic to
${\rm SU}(2)\times{\rm SU}(2)=S^3\times S^3$ as a manifold.  The
diffeomorphism is
\begin{equation}
(g_1,g_2,g_3){\rm SU}(2) \mapsto (g_1g_3^{-1},g_2g_3^{-1})\ .
\end{equation}
The left action of ${\rm SU}(2)\times{\rm SU}(2)\times 1$ on the coset
space coincides with the left action of ${\rm SU}(2)\times{\rm SU}(2)$ on
itself, while the left action of $1\times1\times{\rm SU}(2)$ on the coset
space corresponds to a right action of a diagonal ${\rm SU}(2)$ on
${\rm SU}(2)\times{\rm SU}(2)$.  We denote this latter action ${\rm SU}(2)_R$.  The basic idea is that one can impose symmetry under ${\rm SU}(2)^2$ and ${\rm SU}(2)_R$ separately.

Let $\hat e^\a,\check e^\a$ denote left-invariant 1-forms on two copies
of $S^3=\,$SU(2), normalised so that
\begin{equation}
\diff\hat e^\a=-\frac{1}{2R}\eps_{\a\b\g}\hat e^\b\wedge\hat e^\g\ ,
\end{equation}
and similar for $\check e^\a$.  Both $\hat e^\a$ and $\check e^\a$ transform
in the irreducible 3-dimensional representation of ${\rm SU}(2)_R$.
The nearly K\"ahler structure is determined by setting
\begin{equation}
 \Th^\a = \frac{1}{3}(\tau \hat e^\a +\bar\tau\check e^\a)\ , \qquad
 \tau = -\frac{1}{2}+\im\frac{\sqrt3}{2}\ .
\end{equation}
Note that the corresponding metric is not the usual product metric on
$S^3\times S^3$.  The forms $\Th^\a$ satisfy
\begin{equation}
\diff \Th^\a = -\frac{1}{2R}(2\eps^{\a\bar\b\bar\g}\Th^\b\wedge\Th^\g -
\eps^{\a\b\g}\Th^{\bar\b}\wedge\Th^{\bar\g} +
2\eps^{\a\b\bar\g}\Th^{\bar\b}\wedge\Th^\g )\ .
\end{equation}
The (1,0)-forms $\Th^\a$ define an SU(3)-structure written in the standard form (\ref{SU3 structure}) on $S^3\times S^3$; the easiest way to see that it is nearly K\"ahler is to compute,
\begin{equation}
\diff \omega = \frac{3}{R}{\rm Im}\Omega\ ,\quad \diff\Omega = \frac{2}{R}\omega\wedge\omega\ .
\end{equation}

Let $A$ denote a gauge field on $\R^2$ with gauge group $K$ and let
$\phi_\a$ denote three adjoint scalars.  Then the following ansatz
describes the most general possible SU(2)$^2$-invariant gauge field
on $\R^2\times S^3\times S^3$:
\begin{equation}
\Acal = A + \phi_\a\Th^\a + \phi_{\bar\a}\Th^{\bar\a}\ .
\end{equation}
The Spin(7)-instanton equations for $\Acal$ are reduced to
\begin{eqnarray}
e^{\im\xi} D_{\bar z}\phi_{\bar\a} &=&
\frac{1}{2}\eps_{\bar\a\bar\b\bar\g}[\phi_\b,\phi_\g] +
\frac{1}{R}(\phi_{\bar\a}-2\phi_\a)\ , \\
\ast F &=& 2\im [\phi_\a,\phi_{\bar\a}]\ .
\end{eqnarray}

Let us emphasise again that we have imposed only ${\rm SU}(2)^2$ symmetry, whereas the method described in subsection \ref{sec:gen nk} imposes ${\rm SU}(2)^3$ symmetry.  The additional ${\rm SU}(2)_R$ symmetry can be imposed within the present method, if desired.  For example, with gauge group $K={\rm SU}(2)$, the ansatz,
\begin{equation}
\phi_\a = \frac{\im}{2R}(\phi-1) \sigma_\a\ , \quad A=0\ ,
\end{equation}
with $\phi$ a complex scalar, is invariant under a combination of ${\rm SU}(2)_R$ rotations and gauge transformations.  Then the Spin(7)-instanton equations become
\begin{equation}
\label{eq:8.2.1}
e^{\im\xi}\partial_{\bar z}\bar\phi = \frac{1}{R}(\bar\phi-\phi^2)\ .
\end{equation}
This is the equation that we would have obtained if we had followed the method of subsection \ref{sec:gen nk}.

\subsection{Vortices and invariant instantons}

We have already emphasised that the four nearly K\"ahler coset spaces are algebraically very similar.  One consequence is that the reduced Spin(7)-instanton equations written down in the previous subsections are all related to each other.  Clearly, the equations (\ref{eq:8.1.1}) and (\ref{eq:8.2.1}) arising in the cases $G_2/{\rm SU}(3)$ and ${\rm SU}(2)^3/{\rm SU}(2)$ coincide.  These in turn are obtained from equations (\ref{eq:8.3.1})-(\ref{eq:8.3.3}) arising in the case ${\rm Sp}(2)/{\rm Sp}(1){\times} {\rm U}(1)$, on setting $\phi=\chi,a=0$ and $\Lambda=1$.  Finally, the equations (\ref{eq:8.4.1})-(\ref{eq:8.4.5}) arising in the case ${\rm SU}(3)/{\rm U}(1)^2$ specialise to (\ref{eq:8.3.1})-(\ref{eq:8.3.3}) on setting $\chi_1=\chi$, $\chi_2=\chi_3=\phi$, $b=0$.  As a result, solving any one of the above equations may lead to solutions of some of the others.

The Hermitian flow structure can help in the search for solutions.  We start by discussing equation (\ref{eq:8.1.1}), which is a Hermitian flow for
\begin{equation}
h = \sfrac{3}{2}\diff\phi\diff\bar\phi \ ,\quad W = \sfrac{1}{2R}(3|\phi|^2-\phi^3-\bar\phi^3)\ .
\end{equation}
By making a rotation of $z=x+\im y$ we can fix $\xi=0$; then by imposing $\pa_y=0$ or $\pa_x=0$ the Hermitian flow reduces to a gradient flow or a Hamiltonian flow respectively.  Solutions of these equations on $\R$ were discussed in \cite{HILP}; in the present context these lift to solutions of the Spin(7)-instanton equations on $M^6\times\R\times S^1$, with $M^6$ any nearly K\"ahler coset space.  Solutions of the Hamiltonian flow also exist on $S^1$: these move around the closed trajectories $W=C$ for $C<1/2R$.  In the present context these give us Spin(7)-instantons on $M^6\times T^2$.

Equations (\ref{eq:8.3.1})-(\ref{eq:8.3.3}) with $\Lambda=1$ are also a Hermitian flow, this time for
\begin{equation}
h = \sfrac{1}{2}(2\diff\phi\diff\bar\phi+\diff\chi\diff\bar\chi) \ ,\quad W = \sfrac{1}{2R}(2|\phi|^2+|\chi|^2-\phi^2\chi-\bar\phi^2\bar\chi)\ .
\end{equation}
The Hamiltonian flow preserves not only $W$, but also $U=|\phi|^2-|\chi|^2$.  This is because $U$ is the moment map generating the U(1) gauge symmetry of $W$.  So the Hamiltonian flow is integrable, a fact which should help the search for solutions.  Similar comments apply to equations (\ref{eq:8.4.1})-(\ref{eq:8.4.5}).

Equations (\ref{eq:8.3.1})-(\ref{eq:8.3.3}) also possess solutions with $a\neq0$.  Choose a gauge so that $\phi=e^f$ for some real function $f$, and suppose that $\chi=\bar\phi^2=e^{2f}$.  Then equation (\ref{eq:8.3.2}) implies that $a_{\bar z}=\im\pa_{\bar z}f$.  Equations (\ref{eq:8.3.1}) and (\ref{eq:8.3.3}) become
\begin{eqnarray}
2\pa_{\bar z} f &=& \frac{1}{R}(1-e^{2f})\ , \\
4\pa_z\pa_{\bar z} f &=& \frac{2}{R^2}e^{2f}(e^{2f}-1)\ .
\end{eqnarray}
Surprisingly the second of these equations is implied by the first, so we need only solve the first.  A solution is
\begin{equation}
f = -\frac{1}{2}\ln (1+e^{-2x/R})\ .
\end{equation}
This gives an Spin(7)-instanton on $M^6\times\R\times S^1$, with $M^6={\rm Sp}(2)/{\rm Sp}(1)\times U(1)$ or ${\rm SU}(3)/{\rm U}(1)^2$.

All of the solutions presented above are essentially 1-dimensional; we have not searched for genuine 2-dimensional solutions with $\Lambda=1$.  However, studying the vacua of these equations (i.e. the $z$-independent solutions) can give hints as to what 2-dimensional solutions might exist.  First, equation (\ref{eq:8.1.1}) has a $\mathbb{Z}_3$-symmetry, and its four vacua are $\phi=0$ and $\phi=$ a cube root of unity.  It is quite similar to an equation studied in \cite{junction}, except that the function $W=3|\phi|^2-\phi^3-\bar\phi^3$ playing the role of superpotential is not holomorphic.  Equation (\ref{eq:8.1.1}) might admit domain wall junction solutions, similar to those found in \cite{junction}.

Second, the non-zero vacua of (\ref{eq:8.3.1})-(\ref{eq:8.3.3}) or (\ref{eq:8.4.1})-(\ref{eq:8.4.5}) form topologically non-trivial manifolds.  The vacuum manifolds are $|\phi|^2=1$, $\chi=\bar\phi^2$ and $|\chi_2|^2=|\chi_3|^2=1$, $\chi_1=\bar\chi_2\bar\chi_3$, which are topologically a circle and 2-torus, respectively.  Thus field configurations on $\R^2$ which tend to the vacuum at infinity possess a topological charge, the winding number of the field at infinity.  It would be interesting to see whether topologically non-trivial solutions to these equations exist.

We have been able to find 2-dimensional solutions in the case where $\Lambda>1$.  Setting $\phi=0$ in (\ref{eq:8.3.1})-(\ref{eq:8.3.3}) gives
\begin{eqnarray}
\left(\pa_{\bar z} - 2\im a_{\bar z} - \frac{1}{\Lambda^3R}\right) \bar\chi &=& 0\ , \\
\ast\diff a &=& \frac{2\Lambda^2}{R^2}\left( |\chi|^2 -1+\frac{1}{\Lambda^4}\right)\ .
\end{eqnarray}
For $\Lambda>1$ these are mathematically equivalent to the BPS abelian Higgs vortex equations, with gauge field $\tilde a = a -(1/\Lambda^3R)\diff y$.  Thus solutions of the abelian Higgs vortex equations on $\Sigma^2$ yield Spin(7)-instantons on $M^6\times\Sigma^2$, with $M^6={\rm Sp}(2)/{\rm Sp}(1)\times {\rm U}(1)$ or ${\rm SU}(3)/{\rm U}(1)^2$ and $\Sigma^2=\R^2$, $\R\times S^1$ or $T^2$.

\section{Conclusions}
\label{sec:9}

We have studied the Spin(7)-instanton equations on 8-manifolds of the form $M^d\times G/H$, with $M^d=T^d$ or one of its decompactifications $T^{d-p}\times\R^p$.  By imposing $G$-invariance, the Spin(7)-instanton equations reduce to equations on $M^d$.  Where possible, we have constructed solutions, although our search was not exhaustive and we anticipate that further solutions could be found with more work.  One could also take $M^d$ to be any manifold admitting an SU($n$)-structure for $n\leq2d$; our calculations should go through with only minor changes in this more general situation.

Since all of the compact spaces we considered were elliptic fibrations, Spin(7)-instantons on them are potentially relevant in F-theory compactifications, although compactification of F-theory on non-Calabi-Yau manifolds is far from being understood.  Nevertherless, we hope that our examples have served to illustrate that there is something to be said about generalised instantons, and torsionful Yang-Mills theory, on manifolds with non-vanishing intrinsic torsion.  Of course, the methods that we have used could easily be adapted to studying the $G_2$-instanton and Hermitian-Yang-Mills equations in seven and six dimensions, which can be obtained from the Spin(7)-instanton equations by dimensional reduction.  These equations are relevant to M- and string theory compactifications.

A natural direction for further study would be to solve the supersymmetry
constraint equations of heterotic supergravity by using solutions to all
these instanton equations.  It is also of interest to extend the techniques of
the equivariant dimensional reduction for K\"ahler coset spaces~\cite{LPS}
to higher rank bundles over homogeneous manifolds with SU(4)-structure.

\section*{Acknowledgments}

\noindent
We would like to thank Olaf Lechtenfeld for useful remarks.  DH wishes to thank Nick Manton from bringing \cite{junction} to his attention.  This work was partially supported by the Deutsche Forschungsgemeinschaft
and the Russian Foundation for Basic Research (grants 08-01-00014-a and
09-02-91347) and the Engineering and Physical Sciences Research Council (grant EP/G038775/1).

\bigskip
%\newpage


\begin{thebibliography}{99}

\bibitem{Corrigan:1982th}
  E.~Corrigan, C.~Devchand, D.B.~Fairlie and J.~Nuyts,
  ``First order equations for gauge fields in spaces of dimension 
    greater than four,''
  Nucl.\ Phys.\ B {\bf 214} (1983) 452.

\bibitem{DT}
  S.K.~Donaldson and R.P.~Thomas,
  ``Gauge theory in higher dimensions,''\\
  in: {\it The Geometric Universe},
  Oxford University Press, Oxford, 1998.

\bibitem{Bau}
  L.~Baulieu, H.~Kanno and I.M.~Singer,
  ``Special quantum field theories in eight and other dimensions,''
  Commun.\ Math.\ Phys.\ {\bf 194} (1998) 149
  [arXiv:hep-th/9704167].

\bibitem{Kapustin:2006pk}
  A.~Kapustin and E.~Witten,
  ``Electric-magnetic duality and the geometric Langlands program,''
  arXiv:hep-th/0604151.

\bibitem{Witten:2010cx}
  E.~Witten,
  ``Analytic continuation of Chern-Simons theory,''
  arXiv:1001.2933 [hep-th].

\bibitem{DUY}
  S.K.~Donaldson,
  ``Anti-self-dual Yang-Mills connections on a complex algebraic surface
    and stable vector bundles,''
  Proc.\ Lond.\ Math.\ Soc.\ {\bf 50} (1985) 1;\\
  ``Infinite determinants, stable bundles and curvature,''
  Duke Math.\ J.\ {\bf 54} (1987) 231;

 K.K.~Uhlenbeck and S.-T.~Yau,
  ``On the existence of Hermitian Yang-Mills connections in stable vector bundles,''
  Commun.\ Pure Appl.\ Math.\ {\bf 39} (1986) S257;\\
  ``A note on our previous paper,''
  {\sl ibid.} {\bf 42} (1989) 703.

\bibitem{SaEarp}
  H.N.~S\`a Earp,
  ``Instantons on $G_2$-manifolds'',
  PhD thesis, Imperial College London, 2009.

\bibitem{Lewis}
  C.~Lewis,
  ``Spin(7) instantons'',
  PhD thesis, Oxford University, 1998.

\bibitem{WFN}
  R.S.~Ward,
  ``Completely solvable gauge field equations in dimension
    greater than four,''
  Nucl.\ Phys.\ B {\bf 236} (1984) 381;
  %%CITATION = NUPHA,B236,381;%%
 
D.B.~Fairlie and J.~Nuyts,
  ``Spherically symmetric solutions of gauge theories in eight dimensions,''
  J.\ Phys.\ A {\bf 17} (1984) 2867.
  %%CITATION = JPAGB,A17,2867;%%


\bibitem{FNIP}
  S.~Fubini and H.~Nicolai,
  ``The octonionic instanton,''
  Phys.\ Lett.\ B {\bf 155} (1985) 369;
  %%CITATION = PHLTA,B155,369;%%

  T.A.~Ivanova and A.D.~Popov,
  ``Self-dual Yang-Mills fields in $d{=}7, 8$, octonions and Ward equations,''
  Lett.\ Math.\ Phys.\  {\bf 24} (1992) 85;
  %%CITATION = LMPHD,24,85;%%
  ``(Anti)self-dual gauge fields in dimension $d{\ge}4$,''
  Theor.\ Math.\ Phys.\ {\bf 94} (1993) 225.
  %%CITATION = TMPHA,94,225;%%

\bibitem{DS}
 S.K.~Donaldson and E.~Segal,
  ``Gauge theory in higher dimensions II'',
  arXiv:0902.3239 [math.DG].

\bibitem{P1}
A.D.~Popov,
  ``Hermitian-Yang-Mills equations and pseudo-holomorphic bundles
    on nearly K\"ahler and nearly Calabi-Yau twistor 6-manifolds,''
  Nucl.\ Phys.\  B {\bf 828} (2010) 594
  [arXiv:0907.0106 [hep-th]];
%%CITATION = NUPHA,B828,594;%%
  ``Non-Abelian vortices, super-Yang-Mills theory and Spin(7)-instantons,''
  arXiv:0908.3055 [hep-th].
  %%CITATION = ARXIV:0908.3055;%%

\bibitem{HILP}
D.~Harland, T.A.~Ivanova, O.~Lechtenfeld and A.D.~Popov,
  ``Yang-Mills flows on nearly K\"ahler manifolds and $G_2$-instantons,''
  arXiv:0909.2730 [hep-th].
  %%CITATION = ARXIV:0909.2730;%%

\bibitem{flux}
  M.~Grana,
  ``Flux compactifications in string theory: A comprehensive review,''\\
  Phys.\ Rept.\ {\bf 423} (2006) 91
  [arXiv:hep-th/0509003];
  %%CITATION = PRPLC,423,91;%%

  M.R.~Douglas and S.~Kachru,
  ``Flux compactification,''
  Rev.\ Mod.\ Phys.\ {\bf 79} (2007) 733
  [arXiv:hep-th/0610102];
  %%CITATION = RMPHA,79,733;%%

  R.~Blumenhagen, B.~Kors, D.~L\"ust and S.~Stieberger,
  ``Four-dimensional string compactifications with D-branes, orientifolds
    and fluxes,''
  Phys.\ Rept.\ {\bf 445} (2007) 1
  [arXiv:hep-th/0610327].
  %%CITATION = PRPLC,445,1;%%

\bibitem{Lust:2004ig}
  D.~L\"ust and D.~Tsimpis,
  ``Supersymmetric AdS(4) compactifications of IIA supergravity,''
  JHEP {\bf 02} (2005) 027
  [arXiv:hep-th/0412250].

\bibitem{CMPZ}
  P.~Manousselis, N.~Prezas and G.~Zoupanos,
  ``Supersymmetric compactifications of heterotic strings with fluxes and
  condensates,''
  Nucl.\ Phys.\  B {\bf 739} (2006) 85
  [arXiv:hep-th/0511122];
  %%CITATION = NUPHA,B739,85;%%

  A.~Chatzistavrakidis and G.~Zoupanos,
  ``Dimensional reduction of the heterotic string over nearly-K\"ahler
  manifolds,''
  JHEP {\bf 09} (2009) 077
  [arXiv:0905.2398 [hep-th]].
  %%CITATION = JHEPA,0909,077;%%

\bibitem{Str}
  A.~Strominger,
  ``Superstrings with torsion,''
  Nucl.\ Phys.\ B {\bf 274} (1986) 253.
%%CITATION = NUPHA,B274,253;%%

\bibitem{Lopes Cardoso:2002hd}
  G.~Lopes Cardoso, G.~Curio, G.~Dall'Agata, D.~L\"ust, P.~Manousselis and G.~Zoupanos,
  ``Non-Kaehler string backgrounds and their five torsion classes,''
  Nucl.\ Phys.\  B {\bf 652} (2003) 5
  [arXiv:hep-th/0211118].

\bibitem{WG}
  J.A.~Wolf, {\it Spaces of constant scalar curvature}, McGraw-Hill, New York, 1967;

  J.A.~Wolf and A. Gray, ``Homogeneous spaces defined by Lie group automorphisms I,II,''\\
  J. Diff. Geom. {\bf 2} (1968) 77, 115.

\bibitem{Gray}
  A. Gray, ``Nearly K\"ahler manifolds,'' J. Diff. Geom. {\bf 4} (1970) 283.

\bibitem{But}
  J.-B.~Butruille,
  ``Homogeneous nearly K\"ahler manifolds'',
  arXiv:math/0612655 [math.DG].

\bibitem{CYT}
  P.~Gauduchon,
  ``Hermitian connections and Dirac operators,''
  Bollettino\ U.\ M.\ I. (7) 11-B (1997) 257;
  
  A.~Opfermann and G.~Papadopoulos,
  ``Homogeneous HKT and QKT manifolds,''
  arXiv:math-ph/9807026;

  J.~Gutowski, S.~Ivanov, G.~Papadopoulos,
  ``Deformations of generalized calibrations and compact non-K\"ahler manifolds with vanishing first Chern class,''
  Asian\ J.\ Math\ {\bf 7} (2003) 39 [arXiv:math/0205012[math.DG]];

  G.~Grantcharov,
  ``Geometry of compact complex homogeneous spaces with vanishing first Chern class,''
  arXiv:0905.0040 [math.DG].

\bibitem{Chiossi:2002tw}
  S.~Chiossi and S.~Salamon,
  ``The intrinsic torsion of SU(3) and $G_2$ structures,''
  arXiv:math/0202282.

\bibitem{Tian}
  G.~Tian,
  ``Gauge theory and calibrated geometry,''
  Ann.\ Math.\ {\bf 151} (2000) 193
  [arXiv:math/0010015 [math.DG]].
  %%CITATION = ANMAA,151,193;%%

\bibitem{Kobayashi-Nomizu1}
  S.~Kobayashi and K.~Nomizu,
  {\it Foundations of Differential Geometry}, vol.1, \\
  Interscience Publishers, 1963.

\bibitem{KVMR}
  Yu.~A.~Kubyshin, I.P.~Volobuev, J.M.~Mourao and G.~Rudolph,
 ``Dimensional reduction of gauge theories, spontaneous compactification
 and model building,''
    Lect.\ Notes Phys.\  {\bf 349} (1990) 1.
  %%CITATION = LNPHA,349,1;%%

\bibitem{Zoupanos}
  D.~Kapetanakis and G.~Zoupanos,
  ``Coset space dimensional reduction of gauge theories,''\\
  Phys.\ Rept.\ {\bf 219} (1992) 1.

\bibitem{junction}
  G.~W.~Gibbons and P.~K.~Townsend,
  ``Bogomol’nyi equation for intersecting domain walls,''
  Phys.\ Rev.\ Lett.\ {\bf 83} (1999) 1727;

  P.~M.~Saffin,
  ``Tiling with almost-BPS-invariant domain-wall junctions,''
  Phys.\ Rev.\ Lett.\ {\bf 83} (1999) 4249;

  S.~M.~Carroll, S.~Hellerman and M.~Trodden,
  ``Domain wall junctions are 1/4 BPS states,''
  Phys.\ Rev.\ D {\bf 61} (2000) 065001.

\bibitem{LPS}
O.~Lechtenfeld, A.D.~Popov and R.J.~Szabo,
``Quiver gauge theory and noncommutative vortices,''
  Prog.\ Theor.\ Phys.\ Suppl.\  {\bf 171} (2007) 258
  [arXiv:0706.0979 [hep-th]];
  %%CITATION = PTPSA,171,258;%%
``SU(3)-equivariant quiver gauge theories and nonabelian vortices,''
  JHEP {\bf 08} (2008) 093
  [arXiv:0806.2791 [hep-th]].
  %%CITATION = JHEPA,0808,093;%%

\end{thebibliography}
\end{document}